\documentclass[pre,floats,aps,superscriptaddress,showpacs]{revtex4}

\usepackage{graphicx}
\usepackage{amsmath}
\usepackage{version} 

\DeclareMathOperator{\arccot}{arccot}

\newcommand{\bv}{{\bf v}}
\newcommand{\br}{{\bf r}}
\newcommand{\brh}{{\hat{\bf r}}}

\newcommand{\bs}{\hat{\boldsymbol \sigma}}

\newcommand{\bo}{\hat{\boldsymbol \omega}}
\newcommand{\dd}{\textrm{d}}

\begin{document}

\title{Dynamics of a tracer granular particle as a non-equilibrium Markov
  process.}

\date{\today}

\author{Andrea Puglisi}
\affiliation{Laboratoire de Physique Th\'eorique (CNRS UMR8627), B\^atiment
  210, Universit\'e Paris-Sud, 91405 Orsay cedex, France}

\author{Paolo Visco}
\affiliation{Laboratoire de Physique Th\'eorique (CNRS UMR8627), B\^atiment
  210, Universit\'e Paris-Sud, 91405 Orsay cedex, France}
\affiliation{Laboratoire de Physique Th\'eorique
et Mod\`eles Statistiques (CNRS
UMR 8626), B\^atiment 100, Universit\'e Paris-Sud, 91405 Orsay cedex, France}

\author{Emmanuel Trizac}
\affiliation{Laboratoire de Physique Th\'eorique
et Mod\`eles Statistiques (CNRS
UMR 8626), B\^atiment 100, Universit\'e Paris-Sud, 91405 Orsay cedex, France}

\author{Fr\'ed\'eric van Wijland}
\affiliation{Laboratoire de Physique Th\'eorique (CNRS UMR8627), B\^atiment
  210, Universit\'e Paris-Sud, 91405 Orsay cedex, France}
\affiliation{Laboratoire de Mati\`ere et Syst\`emes Complexes 
(CNRS UMR 7057), Universit\'e
Denis Diderot (Paris VII), 2 place Jussieu, 75251 Paris cedex 05, France}

\begin{abstract}
  The dynamics of a tracer particle in a stationary driven granular gas is
  investigated. We show how to transform the linear Boltzmann equation
  describing the dynamics of the tracer into a master equation for a
  continuous Markov process. The transition rates depend upon the stationary
  velocity distribution of the gas. When the gas has a Gaussian velocity
  probability distribution function (pdf), the stationary velocity pdf of the
  tracer is Gaussian with a lower temperature and satisfies detailed balance
  for any value of the restitution coefficient $\alpha$. As soon as the
  velocity pdf of the gas departs from the Gaussian form, detailed balance is
  violated. This non-equilibrium state can be characterized in terms of a
  Lebowitz-Spohn action functional $W(\tau)$ defined over trajectories of time
  duration $\tau$. We discuss the properties of this functional and of a
  similar functional $\overline{W}(\tau)$ which differs from the first for a
  term which is non-extensive in time.  On the one hand we show that in
  numerical experiments, i.e.  at finite times $\tau$, the two functionals
  have different fluctuations and $\overline{W}$ always satisfies an
  Evans-Searles-like symmetry. On the other hand we cannot observe the
  verification of the Lebowitz-Spohn-Gallavotti-Cohen (LS-GC) relation, which
  is expected for $W(\tau)$ at very large times $\tau$. We give an argument
  for the possible failure of the LS-GC relation in this situation. We also
  suggest practical recipes for measuring $W(\tau)$ and $\overline{W}(\tau)$
  in experiments.
\end{abstract}
\maketitle

\section{Introduction}

When a collection of macroscopic grains (diameter ranging from $10^{-4}$ cm up
to $10^{-1}$ cm) is vigorously shaken, under appropriate conditions of packing
fraction, amplitude and frequency of the vibration, a stationary gaseous
state, usually referred to as ``granular gas'', can be obtained~\cite{intro1}.
Kinetic energy is dissipated into heat during collisions among grains, and
this energetic loss is balanced by the external driving, i.e. the shaking of
the container. In recent years the interest on granular gases has strongly
increased: it has been shown that the whole machinery of kinetic theory can be
applied on simplified but realistic models, leading to theoretical results in
good agreement with experiments~\cite{barrat,dufty_kin}. At the same time a
growingly rich phenomenology has emerged from the laboratory, showing that a
granular gas is a sort of Pandora's box for non equilibrium statistical
mechanics. Just to mention the most famous and well established aspects of
these systems, we recall that a granular gas displays non-Gaussian velocity
behavior, the breakdown of energy equipartition and of thermodynamic
equilibrium, spontaneous breaking of many symmetries (shear instability,
cluster instability, convection instability, etc.)  and a general strong
tendency to reduce the range of scales available to a hydrodynamic
description.  Nevertheless, in a carefully controlled environment, one can
avoid all spatial effects, obtaining a homogeneous granular gas whose main
feature is to be stationary and out of thermodynamic equilibrium, traversed by
an energy current that flows from the external driving into an irreversible
sink constituted by the inelastic collisions among the grains. In this case it
is straightforward to measure a so-called {\em granular temperature}
$T_g=\langle |\mathbf{v}|^2/d \rangle$ (where $d$ is the space dimension)
which is different, in general, from the temperature $T$ characterizing the
external driving. We stress that, in real experiments, it is hard to obtain a
stochastic way of pumping energy into a granular gas: the simplest situation
is a vertically vibrating mechanism, where density gradients arise. In this
paper our interest goes to models where the action of the external driving is
in fact that of a homogeneous thermostat. Such a kind of homogeneous heating
has already been achieved experimentally with a granular mono-layer vibrated
on a rough plate \cite{prevost}.  Very recently granular gases have been used
to probe the latest theoretical results concerning fluctuations in
non-equilibrium systems. In particular the so-called Gallavotti-Cohen
Fluctuation Relation (GCFR)~\cite{gc} has been put under
scrutiny~\cite{feitosa,aumaitre}. This relation is a constraint in the
probability distribution function (pdf) of the fluctuations of entropy flux in
the system, which is rigorously demonstrated under certain hypothesis when the
entropy flux is measured by the phase space contraction rate.  Of course an
all-purpose definition of ``entropy flux'' in systems out of equilibrium does
not exist.  The authors of those recent studies have tried to use the rate of
energy injection from the external driving, i.e. the injected power, as a
entropy flux. Even if in~\cite{feitosa} a verification of the GCFR was
claimed, other studies have shown that the situation is more complex and that,
in general, the injected power in a granular gas cannot satisfy such
relation~\cite{us}. The main difference between a granular gas and the
prototypical system that should obey the GCFR is the invariance under
microscopic time reversal, which is violated by inelastic collisions.

In this paper we have tried to follow a simplified line of reasoning,
obtaining a recipe to measure a ``flux'' which is constructed {\em ad hoc}
with the aim of satisfying the GCFR, or at least its Markovian counterpart,
which has been put forward by Lebowitz and Spohn~\cite{lebowitz}. In the
following we will refer to this relation as to the GC-LS Fluctuation Relation.
Our idea is to bypass the problem of ``strong irreversibility'' posed by
inelastic collisions (i.e. the fact that the probability of observing the time
reversibility of an inelastic collision is strictly zero), focusing on the
evolution of a tracer particle~\cite{martin,dufty_diff}. Indeed, in a dilute gas, a
tracer particle performs a continuous Markov process characterized by
transition rates which are always defined, i.e. for any observable transition
the time-reversed transition has a non-zero probability. Therefore, the
projection of a $N$-body system onto a $1$-body system somehow increases its
degree of time reversibility.

Our original aim was to find a quantity, for granular gases, which, by
construction, verifies the GC-LS Fluctuation Relation. Besides the use
of a single particle functional is still interesting because it can
easily be reproduced in real experiments. Moreover, the fluctuations of this
functional are much stronger than the fluctuations of some other observable
which is averaged over a large number of particles, and this is an evident
advantage in an experimental verification of a Fluctuation Relation. Anyway,
we will show by numerical simulations that this verification poses unexpected
problems, even after having followed carefully the recipe given by Lebowitz
and Spohn.  Only at very large integration times and for situations very far
from equilibrium, the GC-LS Fluctuation Relation is near to be satisfied. At
the same time an alternative functional can be defined, differing from the
first one by an apparently small term. This second functional always satisfies
a relation which we consider the Markovian counterpart of the Evans-Searles
(ES) Fluctuation Relation~\cite{es,es_long}. In summary, our investigation has
led to a concrete example, ready to become a real experiment, where the
difference between two Fluctuation Relations, their meaning and
their limits can directly be probed.

The structure of the paper is as follows: in section II we define our
model and its description in terms of a continuous Markov process,
i.e. giving the transition rates that enter its Master Equation. In
this section we also show that, whenever the surrounding gas has a
non-Gaussian velocity distribution, the tracer particle violates
detailed balance and is therefore out of equilibrium. In section III
we define the action functionals that are expected to verify the GC-LS
and ES Fluctuation Relations. In section IV we discuss the results of
numerical simulations, showing a way to optimally measure the
transition rates and finally discussing the measure of the
fluctuations of the action functionals and their verification of the
Fluctuation Relations. We will draw our conclusions in section V. The
derivation of the Master Equation from the linear Boltzmann Equation,
is given in the appendix.

\section{The master equation.}

We consider the dynamics of a tracer granular particle in a
homogeneous and dilute gas of grains which is driven by an unspecified
energy source. The first essential feature of the gas is its spatial
and temporal homogeneity. The tracer particle collides sequentially
with particles of the gas coming from the same ``population'',
independently of the position and time of the collision. The gas is
characterized by its velocity probability density function
$P(\mathbf{v})$, which in turn is determined by the unspecified
details of the model, such as the properties of the driving or of the
grains. It is well known that the velocity probability density
function of a driven granular gas is non-Gaussian: in a homogeneous
setup the slight departure from Gaussianity is well reproduced by a
first Sonine correction (which will be made explicit in the
following). This is expected theoretically and well
verified in experiments~\cite{kicks}.

The second essential property required is the diluteness of the gas:
it guarantees that the evolution as well as the stationary regime of
the velocity probability density function $P_*(\mathbf{v})$ of the
tracer is governed by a Linear Boltzmann equation. In this section we
give expressions for the tracer transition rates in generic dimension
$d$ and for a generic inter-particle (short range) potential. This
potential is parametrized by a parameter $\gamma$ which is the
exponent of the term $|(\mathbf{v}-\mathbf{V}) \cdot
\bo|^\gamma$ representing the collision kernel in
the Linear Boltzmann equation ($\mathbf v$ and $\mathbf{V}$ are the
colliding velocities while $\bo$ is the unitary
vector joining the particles). For example, a value of $\gamma=1$
corresponds to the Hard Sphere potential (which will be studied in the
rest of the article, without losing generality in the results), a
value of $\gamma=0$ corresponds to the Maxwell Molecules case and a
value of $\gamma=2$ to the so-called Very Hard particles.

The inelastic collisions with the gas particles, which solely
determine the instantaneous changes of the velocity of the tracer, are
described by the simplest and most used inelastic collision rule:
\begin{equation}
\mathbf{v}'=\mathbf{v}-m\frac{1+\alpha}{m+M}[(\mathbf{v}-
\mathbf{V})\cdot \bo] \bo
\end{equation}
where $\mathbf{v}$ and $\mathbf{v}'$ are the velocities of the tracer
before and after the collision respectively, $\mathbf{V}$ is the
velocity of the gas particle, $m$ and $M$ are the masses of the tracer
and of the gas particle respectively, while $\bo$ is
the unitary vector joining the centers of the two particles. In the
following we will assume $m=M$, but this will not change the
generality of our results, in view of the following relation
\begin{equation}
m\frac{1+\alpha}{m+M} \equiv \frac{1+\alpha'}{2}\;,\;\;\;{\text with}\;\;\;\;\alpha'=\frac{m-M+2m\alpha}{m+M}
\end{equation}

The $\gamma$-Boltzmann equation for the tracer, in generic dimension
$d$, reads:
\begin{equation} \label{boltzmann}
\frac{\dd P_*(\mathbf{v},t)}{\dd t}=\frac{v_0^{1-\gamma}}{l_0}\int \dd \mathbf{v}_1 \int \dd \mathbf{v}_2 \int' \dd \bo |(\mathbf{v}_1-\mathbf{v}_2) \cdot \bo |^\gamma
P_*(\mathbf{v}_1) P(\mathbf{v}_2) \left\{\delta\left(\mathbf{v}-\mathbf{v}_1+\frac{1+\alpha}{2}[(\mathbf{v}_1-\mathbf{v}_2)\cdot \bo] \bo\right)
-\delta(\mathbf{v}-\mathbf{v}_1)\right\}
\end{equation}
where the primed integral denotes that the integration is performed on
all angles that satisfy $(\mathbf{v}_1-\mathbf{v}_2) \cdot
\bo >0$. The mean free path $l_0$ appears in front of
the collision integrals. In the $\gamma \neq 1$ cases, to respect
dimensionality, a factor $v_0^{1-\gamma}$ is also present, where $v_0$
is the gas thermal velocity. In the following (when not stated
differently) we will put $l_0=1$ and $v_0=1$, which can be always
obtained by a rescaling of time.

From the analysis of equation~(\ref{boltzmann}) given in Appendix,
we find that the evolution of the velocity probability density
function of the tracer is governed by the following Master equation:

\begin{equation} \label{eq:markov}
\frac{\dd P_*(\mathbf{v},t)}{\dd t}=\int \dd \mathbf{v}_1 P_*(\mathbf{v}_1) K(\mathbf{v}_1, \mathbf{v}) - \int \dd \mathbf{v}_1 P_*(\mathbf{v}) K(\mathbf{v}, \mathbf{v}_1).
\end{equation}
where $P_*(v)$ is the velocity pdf of the test particle.
The transition rate $K(\mathbf{v}, \mathbf{v}')$ of jumping from
$\mathbf{v}$ to $\mathbf{v}'$ is given by the following formula:
\begin{equation}
\label{eq:K}
K(\bv,\bv')=\left(\frac{2}{1+\alpha}\right)^{\gamma+1}
\frac{v_0^{1-\gamma}}{l_0} |\Delta \bv|^{\gamma-d+1} 
\int \dd \bv_{2 \tau} P[\mathbf{v}_2(\mathbf{v},\mathbf{v}',\bv_{2 \tau})],
\end{equation}
where $\Delta \bv=\bv' - \bv$ denotes the change of velocity of the
test particle after a collision. The vectorial function $\bv_2$ is defined as
\begin{equation}
\mathbf{v}_2(\mathbf{v},\mathbf{v}',\bv_{2\tau})=v_{2\sigma}
(\mathbf{v},\mathbf{v}')\bs(\mathbf{v},\mathbf{v}')+
 \bv_{2\tau}  \label{eq:Kb},
\end{equation}
where $\bs(\mathbf{v},\mathbf{v}')$ is the unitary vector parallel to $\Delta
\bv$, while $\bv_{2 \tau}$ is entirely contained in the $(d-1)$-dimensional
space perpendicular to $\Delta \bv$ (i.e.  $\bv_{2 \tau} \cdot \Delta \bv
=0$).  This implies that the integral in expression~\eqref{eq:K} is
$(d-1)$-dimensional. Finally, to fully determine the transition rate
(\ref{eq:K}), the expression of $v_{2 \sigma}$ is needed:
\begin{equation}
v_{2 \sigma}(\bv , \bv') = \frac{2}{1+ \alpha} |\Delta \bv| 
+ \bv \cdot \bs \,\,\,.
\end{equation}

Note that equation~(\ref{eq:markov}) is a Master equation for the
evolution of a probability density function. We simplify the
terminology, using the name ``transition rate'' for the function
$K(\mathbf{v}, \mathbf{v}')$, which actually is a ``transition rate
density''. This density has its natural definition in the following
limit:

\begin{equation} \label{definition}
K(\mathbf{v},\mathbf{v}')=\lim_{|\dd\mathbf{v}'| \to 0}
\frac{\mathcal{K}(\mathbf{v} \to \mathbf{u}\in 
\mathcal{B}_{\dd\mathbf{v}'}(\mathbf{v}'))}{|\dd\mathbf{v}'|}
\end{equation}
where $\mathcal{K}(\mathbf{v} \to \mathbf{u}\in
\mathcal{B}_{\dd\mathbf{v}'}(\mathbf{v}'))$ is the probability that the
tracer, after a collision, has a velocity $\mathbf{u}$ contained in a
sphere $\mathcal{B}_{\dd\mathbf{v}'}(\mathbf{v}')$ of radius
$\dd\mathbf{v}'$ centered in the vector $\mathbf{v}'$, having a velocity
$\mathbf{v}$ before the collision.

\subsection{Examples: Gaussian and first Sonine correction}

If $P(\mathbf{v})=\frac{1}{(2\pi T)^{d/2}}
\exp\left({-\frac{v^2}{2T}}\right)$, it then immediately follows that
the transition rate $K(\bv, \bv')$ reads:
\begin{equation} \label{K_gauss}
K(\mathbf{v},\mathbf{v}')=\left(\frac{2}{1+\alpha}\right)^{\gamma+1}
\frac{v_0^{1-\gamma}}{l_0} \left\lvert\Delta v\right\lvert^{\gamma-d+1}
\frac{1}{\sqrt{2\pi T}} e^{- \frac{v_{2 \sigma}^2}{2 T}}\,\,\,.
\end{equation}

In kinetic theory, one of the most used corrections to the Gaussian is the
first non-zero Sonine polynomial approximation~\cite{kicks,sonine}. This means
assuming that $P(\mathbf{v})=\frac{1}{(2\pi T)^{d/2}}\exp{ \left( -
    \frac{v^2}{2T}\right)}(1+a_2S_2^{d}(v^2/2T))$ with $S_2^{d}(x)=\frac{1}{2}
x^2- \frac{d+2}{2} x+\frac{d(d+2)}{8}$. 
The calculation of the integral needed to have an explicit expression
of the transition rate is straightforward:
\begin{equation}
\int
\dd \bv_{2\tau}P(\mathbf{v}_2)=\frac{e^{-\frac{v_{2\sigma}^2}{2T}}}
{\sqrt{2 \pi T}}\left(1+a_2S_2^{d=1}(v_{2\sigma}^2/2T)\right)\,\,\,.
\end{equation}
This leads to
\begin{equation} \label{K_nongauss}
K(\mathbf{v},\mathbf{v}')= \left(\frac{2}{1+\alpha}\right)^{\gamma+1}\frac{v_0^{1-\gamma}}{l_0} \left\lvert\Delta v\right\lvert^{\gamma-d+1}
\frac{1}{\sqrt{2\pi T}}e^{-\frac{v_{2 \sigma}^2}{2T}}\left(1+a_2S_2^{d=1}\left(\frac{v_{2\sigma}^2}{2T} \right) \right)
\end{equation}

\subsection{Detailed balance}

Here, we obtain a simple expression for the ratio between
$K(\mathbf{v},\mathbf{v}')$ and $K(\mathbf{v}',\mathbf{v})$. When
exchanging $\mathbf{v}$ with $\mathbf{v}'$ the unitary vector $\bs$ change of sign.
Furthermore one has that $v_{2 \sigma}(\bv, \bv') \ne v_{2 \sigma}(\bv', \bv)$.
Finally it must be recognized that 
\begin{equation}
\frac{K(\mathbf{v},\mathbf{v}')\dd \mathbf{v}'}{K(\mathbf{v}',\mathbf{v})\dd\mathbf{v}}=
\frac{K(\mathbf{v},\mathbf{v}')}{K(\mathbf{v}',\mathbf{v})}.
\end{equation}
This equivalence is naturally achieved by considering the
definition~(\ref{definition}) and taking uniformly the two limits
$|\dd\mathbf{v}| \to 0$ and $|\dd\mathbf{v}'| \to 0$.
From all these considerations and from equation~\eqref{eq:K} one obtains
immediately:
\begin{equation}
\frac{K(\mathbf{v},\mathbf{v}')}{K(\mathbf{v}',\mathbf{v})}=
\frac{\int \dd \bv_{2\tau}P[\mathbf{v}_2(\bv, \bv')]}{\int \dd \bv_{2\tau}P[\mathbf{v}_2(\bv', \bv)]}
\equiv  \frac{P[v_{2 \sigma} (\bv, \bv')]}{P[v_{2 \sigma} (\bv', \bv)]}\,\,\,.
\end{equation}
We note that this ratio depends only on the choice of the pdf of the gas, $P$, and not
on the other parameters (such as $\gamma$ or $\alpha$). However in realistic
situations (experiments or Molecular Dynamics simulations) $P$ is not a free
parameter but is determined by the choice of the setup (e.g. external
driving, material details, geometry of the container, etc.).

Introducing the short-hand notation $v_{2 \sigma}=v_{2 \sigma}(\bv, \bv')$, $v_{2 \sigma}'=v_{2 \sigma}(\bv', \bv)$ and $v_{\sigma}^{( \prime )}=\bv^{( \prime )} \cdot \bs$,
we also note that

\begin{equation}
(v_{2\sigma}')^2=v_{2\sigma}^2+(v_{\sigma}+v_{\sigma}')^2-2v_{2\sigma}(v_{\sigma}+v_{\sigma}')\,\,,
\end{equation}
from which it follows that 

\begin{equation} \label{eq:delta2}
\Delta_2 =(v_{2\sigma})^2-(v_{2\sigma}')^2=-\Delta-2\frac{1-\alpha}{1+\alpha}\Delta=-\frac{3-\alpha}{1+\alpha}\Delta\,\,,
\end{equation}
where $\Delta=v_{\sigma}^2-(v_{\sigma}')^2 \equiv
|v|^2-|v'|^2$, i.e. the kinetic energy lost by the test-particle
during one collision. When $\alpha=1$ then $\Delta_2=
-\Delta$ (energy conservation).
From the above considerations it follows that 

\begin{itemize}

\item
in the {\bf Gaussian} case, it is found

\begin{equation}
\log\frac{K(\mathbf{v},\mathbf{v}')}{K(\mathbf{v}',\mathbf{v})}=\frac{\Delta}{2T} +2 \frac{1-\alpha}{1+\alpha}\frac{\Delta}{2T}=\frac{3-\alpha}{1+\alpha}\frac{\Delta}{2T}
\end{equation}

\item
in the {\bf First Sonine Correction} case, it is found

\begin{equation} \label{ratio_sonine}
\log\frac{K(\mathbf{v},\mathbf{v}')}{K(\mathbf{v}',\mathbf{v})}=\frac{3-\alpha}{1+\alpha}\frac{\Delta}{2T}+\log\frac{\left\{1+a_2S_2^{d=1}\left[\frac{\left(\frac{2}{1+\alpha}(v_{\sigma}'-v_{\sigma})+v_{\sigma}\right)^2}{2T}\right]\right\}}{\left\{1+a_2S_2^{d=1}\left[\frac{\left(\frac{2}{1+\alpha}(v_{\sigma}-v_{\sigma}')+v_{\sigma}'\right)^2}{2T}\right]\right\}}
\end{equation}

\end{itemize}

In the case where $P(v)$ is a Gaussian with temperature $T$, it is immediate
to observe that
\begin{equation}
P_*(\mathbf{v})K(\mathbf{v},\mathbf{v}')=
P_*(\mathbf{v}')K(\mathbf{v}',\mathbf{v})
\end{equation}
if $P_*$ is equal to a Gaussian with temperature
$T'=\frac{\alpha+1}{3-\alpha}T \le T$. This means that there is a Gaussian
stationary solution of equation~\eqref{eq:markov} (in the Gaussian-bulk case),
which satisfies detailed balance. The fact that such a Gaussian with a
different temperature $T'$ is an exact stationary solution was known
from~\cite{martin}. It thus turns out that detailed balance is
satisfied, even out of thermal equilibrium. Of course this is an artifact of
such a model: it is highly unrealistic that a granular gas yields a Gaussian
velocity pdf. As soon as the gas velocity pdf $P(v)$ ceases to be Gaussian,
detailed balance is violated, i.e. the stationary process performed by the
tracer particle is no more in equilibrium within the thermostatting gas.  We
will see in section~\ref{noneq} how to characterize this departure from
equilibrium.

\subsection{Collision rates}

The velocity dependent collision rate is defined as

\begin{equation}
r(\mathbf{v})=\int \dd \mathbf{v}' K(\mathbf{v},\mathbf{v}') 
\equiv \frac{1}{l_0} \int \dd \mathbf{v}' \dd \hat{\mathbf{\sigma}} 
\Theta[(\mathbf{v}-\mathbf{v}') \cdot \hat{\mathbf{\sigma}}] 
(\mathbf{v}-\mathbf{v}')\cdot\hat{\mathbf{\sigma}}P(\mathbf{v}')
\end{equation}
where the last passage is true for hard spheres.
In the following, for simplicity and for direct comparison with numerical results,
we will consider only the model of a two-dimensional hard sphere gas.
In this particular case, the collision rate reads:

\begin{itemize}

\item for a Gaussian bulk:

\begin{equation} \label{gauss_v_rate}
r(v)=\frac{\sqrt{\pi}}{l_0\sqrt{2T}} e^{-\frac{v^2}{4T}}
\left[(2T+v^2)I_0(\frac{v^2}{4T})+v^2I_1(\frac{v^2}{4T})\right]
\end{equation}
where $I_n(x)$ is the $n$-th modified Bessel function of the first kind. Note
that from equation~\eqref{gauss_v_rate} one obtains the total collision
frequency:
\begin{equation}
\omega_c=\frac{1}{T'}\int \dd v v e^{-\frac{v^2}{2T'}} r(v)
=\frac{\sqrt{2\pi (T+T')}}{l_0}
=\frac{2}{l_0}\sqrt{\frac{2\pi}{3-\alpha}T}
\end{equation}
which, using $l_0=\frac{1}{n\sigma^2}$ for the mean free path gives
$\omega_c=2\sqrt{\pi T} n \sigma^2$ in the case $\alpha=1$, i.e. the
known result from kinetic theory for the elastic hard disks. Finally
note that $r(v)$ for the Gaussian case {\em does not depend} upon the
restitution coefficient (while the transition rates and the total
collision frequency do).

\item for a bulk with a Gaussian distribution plus the first Sonine
  approximation:

\begin{equation}
r(\mathbf{v})=\frac{\sqrt{\pi}}{l_08\sqrt{2}T^{5/2}}
e^{-\frac{v^2}{4T}}(8(1+a_2)T^2-8a_2Tv^2+a_2v^4)
\left[ \left(2T+v^2 \right)I_0 \left( \frac{v^2}{4T} \right)
+v^2I_1 \left(\frac{v^2}{4T} \right) \right].
\end{equation}
In this case the expression for $\omega_c$ is more involved.

\end{itemize}

\section{Non-equilibrium characterization} \label{noneq}

From the previous section we have learnt that the dynamics of the velocity of
a tracer particle immersed in a granular gas is equivalent to a Markov process
with well defined transition rates. This means that the velocity of the tracer
particle stays in a state $\mathbf{v}$ for a random time $t\geq 0$ distributed
with the law $r(v) e^{-r(v)t} \dd t$ and then makes a transition to a new value
$\mathbf{v}'$ with a probability $r(v)^{-1}K(\mathbf{v},\mathbf{v}')$. At this
point it is interesting to ask about some characterization of the
non-equilibrium dynamics, i.e. of the violation of detailed balance, which we
know to happen whenever the surrounding granular gas has a non-Gaussian
distribution of velocity.

To this extent, we define two different action functionals,
following~\cite{lebowitz}:

\begin{subequations} \label{entropy}
\begin{align}
  W(t)= \sum_{i=1}^{n(t)} \log \frac{K(\mathbf{v}_i \to
    \mathbf{v}_i')}{K(\mathbf{v}_i' \to \mathbf{v}_i)}\\
  \overline{W}(t)= \log\frac{P_*(\mathbf{v}_1)}{P_*(\mathbf{v}_{n(t)}')}+
  \sum_{i=1}^{n(t)} \log \frac{K(\mathbf{v}_i \to
    \mathbf{v}_i')}{K(\mathbf{v}_i' \to
    \mathbf{v}_i)}\equiv\log\frac{\mathcal{P}(\mathbf{v}_1 \to \mathbf{v}_2
    \to ... \to \mathbf{v}_{n(t)})}{\mathcal{P}(\mathbf{v}_{n(t)} \to
    \mathbf{v}_{n(t)-1} \to ... \to \mathbf{v}_1)} \label{entropyb}
\end{align}
\end{subequations}
where $i$ is the index of collision suffered by the tagged particle,
$\mathbf{v}_i$ is the velocity of the particle before the $i$-th collision,
$\mathbf{v}_i'$ is its post-collisional velocity, $n(t)$ is the total number
of collisions in the trajectory from time $0$ up to time $t$, and $K$ is the
transition rate of the jump due to the collision. Finally, we have used the
notation $\mathcal{P}(\mathbf{v}_1 \to \mathbf{v}_2 \to ... \to \mathbf{v}_n)$
to identify the probability of observing the trajectory $\mathbf{v}_1 \to
\mathbf{v}_2 \to ... \to \mathbf{v}_n$. The quantities $W(t)$ and
$\overline{W}(t)$ are different for each different trajectory (i.e. sequence
of jumps) of the tagged particle. The first term
$\log\frac{P_*(\mathbf{v}_1)}{P_*(\mathbf{v}_{n(t)}')}$ in the definition of
$\overline{W}(t)$, eq.~\eqref{entropyb}, will be called in the following
``boundary term'' to stress its non-extensivity in time.  The two above
functionals have the following properties:

\begin{itemize}
  
\item $W(t) \equiv 0$ if there is exact symmetry, i.e. if $K(\mathbf{v}_i \to
  \mathbf{v}_{i+1})=K(\mathbf{v}_{i+1} \to \mathbf{v}_i)$ (e.g. in the
  microcanonical ensemble); $\overline{W}(t) \equiv 0$ if there is detailed
  balance (e.g. any equilibrium ensemble);
  
\item we expect that, for $t$ large enough, for almost all the trajectories
  $\lim_{s \to \infty} W(s)/s = \lim_{s \to \infty} \overline{W}(s)/s=\langle
  W(t)/t \rangle=\langle \overline{W}(t)/t \rangle$; here (since the system
  under investigation is ergodic and stationary) the meaning of $\langle
  \rangle$ is intuitively an average over many independent segments of a
  single very long trajectory;
 
\item for large enough $t$: 1) at equilibrium $\langle W(t) \rangle=\langle
  \overline{W}(t)\rangle=0$; 2) out of equilibrium (i.e. if detailed balance
  is not satisfied) those two averages are positive; we use those
  equivalent averages, at large $t$, to characterize the distance from
  equilibrium of the stationary system;

\item if $S(t)=-\int \dd v P_*(v,t)\log P_*(v,t)$ is the entropy associated
  to the pdf of the velocity of the tagged particle $P_*(v,t)$ at time $t$ 
  (e.g. $-H$ where $H$ is the Boltzmann-H function), then

\begin{equation}
\frac{\dd}{\dd t}S(t)=R(t)-A(t)
\end{equation}

where $R(t)$ is always non-negative, $A(t)$ is
linear with respect to $P_*$ and, finally, $\langle W(t) \rangle
\equiv \int_0^t \dd t' A(t')$. This leads to consider $W(t)$ equivalent to the
contribution of a single trajectory to the total entropy flux. In a stationary
state $A(t)=R(t)$ and therefore the flux is equivalent to the production; this
property has been recognized in~\cite{lebowitz}.

\item {\bf $FR_W$} (Lebowitz-Spohn-Gallavotti-Cohen fluctuation relation):
  $\pi(w)-\pi(-w)=w$ where $\pi(w)=\lim_{t\to \infty}\frac{1}{t}\log{f_W^t(t
    w)}$ and $f_W^t(x)$ is the probability density function of finding
  $W(t)=x$ at time $t$; at equilibrium the $FR_W$ has no content; note that in
  principle $\pi'(w,t)=\frac{1}{t}\log{f_W^t(t w)}\neq\pi(w)$ at any finite
  time; a generic derivation of this property has been obtained
  in~\cite{lebowitz}, while a rigorous proof with more restrictive hypothesis
  is in~\cite{maes}; the discussion for the case of a Langevin equation is
  in~\cite{kurchan}.
  
\item {\bf $FR_{\overline{W}}$} (Evans-Searles fluctuation relation):
  $\overline{\pi}(w,t)-\overline{\pi}(-w,t)=w$ where
  $\overline{\pi}(w,t)=\frac{1}{t}\log{f_{\overline{W}}^t(t w)}$ and
  $f_{\overline{W}}^t(x)$ is the probability density function of finding
  $\overline{W}(t)=x$ at time $t$; at equilibrium the $FR_{\overline{W}}$ has
  no content;

\end{itemize}

On the one side we have called $FR_{W}$ a ``Lebowitz-Spohn-Gallavotti-Cohen''
fluctuation relation, following~\cite{lebowitz}, where the analogy with the
original Gallavotti-Cohen Fluctuation Relation has been stated explicitly.  On
the other side we have called the $FR_{\overline{W}}$ a ``Evans-Searles''
Fluctuation Relation, inspired by the following analogy. The original
Gallavotti-Cohen Fluctuation Theorem concerns the fluctuation of a functional
$\Omega(t)$ which is the time-averaged phase space contraction rate, i.e.
\begin{equation}
t\Omega(t) \equiv \int_0^t\Lambda(\Gamma(s)) \dd s
\end{equation}
where $\Lambda(\Gamma(s))$ is the phase space contraction rate at the point
$\Gamma(s)$ in the phase space visited by the system at time $s$.  On the
other side, the original Evans-Searles Fluctuation Theorem~\cite{es,es_long}
concerns the fluctuation of a functional $\overline{\Omega}(t)$ (often called
``dissipation function'') defined as
\begin{equation} \label{es_orig}
t\overline{\Omega}(t) \equiv
\log\left(\frac{f(\Gamma(0))}{f(\Gamma(t))}\right)-\int_0^t\Lambda(\Gamma(s)) \dd s
\equiv \log \frac{p(\delta V_\Gamma(\Gamma(0))} 
{p(\delta V_\Gamma(I_T\Gamma(t)))}
\end{equation}
where $f(\Gamma)$ is the phase space density function at time $0$,
$p(\delta V_\Gamma(\Gamma))$ is the probability of observing at time
$0$ an infinitesimal phase space volume of size $\delta V_\Gamma$
around the point $\Gamma$, and $I_T$ is the time reversal operator
which typically leaves the position unaltered and changes the sign of
the velocities. The last equality in equation~\eqref{es_orig} follows
from the Liouville equation written in Lagrangian form: $\frac{\dd
f(\Gamma,t)}{\dd t}=-\Lambda(\Gamma)f(\Gamma,t)$ (details on its
derivation can be found in~\cite{es_long}). The ratio of probabilities
appearing at the end of the r.h.s. of~\eqref{es_orig} can be thought
as the deterministic equivalent of the ratio of probabilities
appearing at the end of the r.h.s of definition~\eqref{entropyb}. The
term $\log\frac{f(\Gamma(0))}{f(\Gamma(t))}$ is instead analogous to
the term $\log\frac{P_*(\mathbf{v}_1)}{P_*(\mathbf{v}_{n(t)}')}$,
since the system can be prepared at time $0$ in such a way that its
phase space density function $f$ is arbitrarily near to its stationary
invariant measure. The analogy between those two terms, and therefore
between the two pairs of functionals $\Omega,\overline{\Omega}$ and
$W,\overline{W}$, cannot be derived mathematically rigorously, at the
moment, but formally at least, it looks rather convincing.
Note also that the difference between the two functionals of each
couple vanishes when the invariant phase space distribution function
becomes constant, e.g. in the microcanonical ensemble.

As a final remark of this section, consider the more general case where the
tracer particle feels the external driving too: this means that other
transitions are possible. The simplest case corresponds to an external driving
modeled as a thermostat which acts on the grains in the form of independent
random kicks~\cite{kicks}. It is straightforward to recognize that the
transition rates competing to the velocity changes due to these kicks are
symmetric and therefore do not contribute to the action functionals in
eq.~\eqref{entropy}. On the other hand, when one wants to include the effect
of walls (both still or vibrating), new transition rates should be taken into
account. A collision with a wall can be modeled as a collision with a flat
body with infinite mass, so that
\begin{equation}
v'_\sigma=-\alpha_w v_\sigma + (1+\alpha_w)v_w(t)
\end{equation}
where $\sigma$ is the direction perpendicular to the wall and $v_w(t)$ is the
wall velocity in that direction at the time $t$ of the collision, while
$\alpha_w$ is the restitution coefficient which describes the energy
dissipation during such collision. If the wall is still ($v_w=0)$ and
$\alpha_w<1$ then the transition is non-reversible and this results in a
divergent contribution to the action functionals. If the wall is still but
elastic, then again the transition rates are symmetric and do not contribute
to the action functionals. Finally, if the wall is vibrating the transition
rates will depend upon the probability $P_w(v_w=x)$ of finding the vibrating
wall at a certain velocity $x$.

\section{Numerical simulations}

A Direct Simulation Monte Carlo~\cite{bird} is devised to simulate in
dimension $d=2$ the dynamics of a tracer particle undergoing inelastic
collisions with a gas of particles in a stationary state with a given velocity
pdf  $P(\mathbf{v})$.  The simulated model contains three parameters: the
restitution coefficient $\alpha$, the temperature of the gas $T$ (which we
take as unity) and the coefficient of the first Sonine correction which
parametrizes the velocity pdf of the gas, $a_2$. The tracer particle has a
``temperature'' $T_*<T$ and a velocity pdf that is observed to be well
described again by a first Sonine correction to a Gaussian, parametrized by a
coefficient $a_2^*$. We also use $l_0=1$. This means that the elastic mean
free time is $\tau_c^{el}=1/2\sqrt{\pi}$ (it is larger in inelastic cases).

The quantities $W(\tau)$ and $\overline{W}(\tau)$ are measured along
independent (non-overlapping) segments of time-length $\tau$ extracted from a
unique trajectory.

\subsection{Verification of transition rates formula}

In this subsection we verify the correctness of formula~\eqref{K_gauss} for
the Gaussian bulk and formula~\eqref{K_nongauss} for the non-Gaussian
(Sonine-corrected) bulk. Since the algorithm used (the DSMC~\cite{bird}) is
known to very well reproduce the Molecular Chaos assumption, which is the
unique hypothesis used to obtain those formula, we consider this verification
a mere check of the consistency of algorithms and calculations. On the other
side, while performing this verification, we have devised a way of optimizing
the measurement of the transition rates, which can be adopted in experiments.

The brute force method of measuring $K$ in an experiment or a simulation is to
observe subsequent collisions for a long time $t$ and cumulate them in a
$4$-dimensional ($2d$-dimensional) matrix $DY_t$: this means that
$DY_t(v_x,v_y,v_x',v_y')$ contains the number of observed collisions such that
the precollisional and postcollisional velocities are contained in a cubic
region centered in $(v_x,v_y,v_x',v_y')$ and of volume $\dd v_x \times \dd v_y
\times \dd v_x' \times \dd v_y'$ finite but small. At the same time it is necessary
to measure $P_*(v_x,v_y) \dd v_x \dd v_y$ for the tagged particle. These quantities
are related by the following formula:
\begin{equation} \label{recipe1}
DY_t(v_x,v_y,v_x',v_y')=tP_*(v_{x},v_{y})K(\mathbf{v}, \mathbf{v}')
\dd v_x \dd v_y \dd v_x' \dd v_y',
\end{equation}
which can be immediately inverted to obtain $K(\mathbf{v},\mathbf{v}')$.
Anyway this recipe may require a very large statistics, because of the high
dimensionality of the histogram $DY_t$.

Then we notice that $K$ is a function of only $v_{\sigma},v_{\sigma'}$. It
becomes tempting, therefore, to reduce the number of variables from $2d$ to
$2$ in order to optimize the procedure. A histogram $D\tilde{Y}_t$ of
dimensionality $2$ (in any dimension) can be obtained, so that each element
$D\tilde{Y}_t(u_\sigma,u_\sigma')$ contains the number of observed collisions
such that the projection along the direction $\hat\sigma$ of the
precollisional velocity is $u_\sigma$ and that of the postcollisional velocity
is $u_\sigma'$. We note that
\begin{multline}
  D\tilde{Y}_t(u_\sigma,u_\sigma')= \dd u_\sigma \dd u_\sigma' \int
  DY_t(v_x,v_y,v_x',v_y')\delta(u_\sigma-\mathbf{v}
  \cdot\hat{\mathbf{\sigma}}(\mathbf{v},\mathbf{v}'))
  \delta(u_\sigma'-\mathbf{v}'\cdot\hat{\mathbf{\sigma}}
  (\mathbf{v},\mathbf{v}'))=\\
  \dd u_\sigma \dd u_\sigma' t \int \dd v_x \dd v_y \dd v_x' \dd v_y' K(\mathbf{v},
  \mathbf{v}')P_*(\mathbf{v}) \delta(u_\sigma-\mathbf{v}
  \cdot\hat{\mathbf{\sigma}}(\mathbf{v},\mathbf{v}'))
  \delta(u_\sigma'-\mathbf{v}'\cdot\hat{\mathbf{\sigma}}
  (\mathbf{v},\mathbf{v}'))=\\
  \dd u_\sigma \dd u_\sigma' t \int \dd v_x \dd v_y \int
  \dd v_\sigma' \dd v_\sigma'\mathcal{J}(v_x,v_y,v_\sigma,v_\sigma')K(\mathbf{v},
  \mathbf{v}')P_*(\mathbf{v})\delta(u_\sigma-v_\sigma)
  \delta(u_\sigma'-v_\sigma').
\end{multline}
In the last passage, we have done the change of variables (at fixed $v_x,v_y$)
$v_x',v_y' \to v_\sigma(v_x,v_y),v_\sigma'(v_x,v_y)$. This implies the
appearing of the associated Jacobian. Moreover, it happens that $\mathbf{v}
\cdot\hat{\mathbf{\sigma}}(\mathbf{v},\mathbf{v}')\equiv v_\sigma$ and
$\mathbf{v}' \cdot\hat{\mathbf{\sigma}}(\mathbf{v},\mathbf{v}')\equiv
v_\sigma'$. Postponing the problem of finding the Jacobian $\mathcal{J}$, we
can absorb the Dirac deltas, obtaining:
\begin{equation}
D\tilde{Y}_t(u_\sigma,u_\sigma')=\dd u_\sigma \dd u_\sigma' t 
\tilde{K}(u_\sigma,u_\sigma')\int \dd v_x \dd v_y 
\mathcal{J}(v_x,v_y,u_\sigma,u_\sigma')P_*(\mathbf{v})
\end{equation}
where $\tilde{K}(u_\sigma,u_\sigma')$ is just the transition probability $K$
as a function of $u_\sigma,u_\sigma'$ (which is its natural representation).
The change of variables is given by the following rule:

\begin{subequations}
\begin{align}
  v_\sigma&=v_x \frac{v_x'-v_x}{\sqrt{(v_x'-v_x)^2+(v_y'-v_y)^2}}
  + v_y \frac{v_y'-v_y}{\sqrt{(v_x'-v_x)^2+(v_y'-v_y)^2}}\\
  v_\sigma'&=v_x' \frac{v_x'-v_x}{\sqrt{(v_x'-v_x)^2+(v_y'-v_y)^2}} + v_y'
  \frac{v_y'-v_y}{\sqrt{(v_x'-v_x)^2+(v_y'-v_y)^2}}.
\end{align}
\end{subequations}
After calculations, the Jacobian reads:
\begin{equation}
\mathcal{J}=\frac{|v_\sigma'-v_\sigma|}
{\sqrt{v_x^2+v_y^2-v_\sigma^2}}\theta(v_x^2+v_y^2-v_\sigma^2).
\end{equation}
From the expression of the Jacobian, and taking into account the isotropy of
$P_*(\mathbf{v})d\mathbf{v}\equiv v\tilde{P}_*(v) \dd v \dd \phi$ one gets
\begin{equation} \label{final}
D\tilde{Y}_t(u_\sigma,u_\sigma')=\dd u_\sigma \dd u_\sigma' t 
\tilde{K}(u_\sigma,u_\sigma')|u_\sigma'-u_\sigma|\mathcal{L}(u_\sigma),
\end{equation}
where
\begin{align} \nonumber \label{prefactor}
  \mathcal{L}(u_\sigma)=\frac{1}{|u_\sigma'-u_\sigma|}\int \dd v_x \dd v_y
  \mathcal{J}(v_x,v_y,u_\sigma,u_\sigma')P_*(\mathbf{v})&= \int \dd v_x \dd v_y
  \frac{P_*(\mathbf{v})}
  {\sqrt{v_x^2+v_y^2-u_\sigma^2}}\theta(v_x^2+v_y^2-u_\sigma^2)=\\
  2\pi\int_{|u_\sigma|}^\infty \dd v v
  \frac{\tilde{P}_*(v)}{\sqrt{v^2-u_\sigma^2}}&= 2\pi\int_0^\infty \dd z
  \tilde{P}_*(\sqrt{z^2+u_\sigma^2}).
\end{align}

As already discussed, the result by Martin and Piasecki~\cite{martin} explains
that when the velocity-pdf of the gas is Gaussian, then also the tagged
particle has a Gaussian pdf. In particular if the bulk has a temperature $T$,
the tagged particle has a temperature $T'=\frac{\alpha+1}{3-\alpha}T$. Then it
is easy to find that
\begin{equation}
K(u_\sigma,u_\sigma')=\left(\frac{2}{1+\alpha}\right)^2
\frac{1}{\sqrt{2 \pi T}}e^{-\frac{1}{2T}
\left(\frac{2}{1+\alpha}(u'_\sigma-u_\sigma)+u_\sigma\right)^2}
\end{equation}
and
\begin{equation}
\mathcal{L}(u_\sigma)=\frac{\sqrt{\pi}}{\sqrt{2T'}}e^{-\frac{u_\sigma^2}{2T'}},
\end{equation}
so that the theoretical expectation for the measured array $D\tilde{Y}_t$ is
\begin{equation} \label{eq:gauss}
D\tilde{Y}_t(u_\sigma,u_\sigma')=\dd u_\sigma \dd u_\sigma' t 
\left(\frac{2}{1+\alpha}\right)^2\frac{1}{2\sqrt{T T'}}|u_\sigma'-u_\sigma|
\exp\left[-\frac{1}{2T}\frac{4}{(1+\alpha)^2}\left(u_\sigma^2+(u_\sigma')^2
+(\alpha-1)u_\sigma u_\sigma' \right)\right]
\end{equation}
which for the case $\alpha=1$ reads
\begin{equation} \label{eq:gauss1}
D\tilde{Y}_t(u_\sigma,u_\sigma')=\dd u_\sigma \dd u_\sigma' t 
\frac{1}{2}\frac{1}{T}|u_\sigma'-u_\sigma|
\exp\left[-\frac{1}{2T}\left(u_\sigma^2+(u_\sigma')^2 \right)\right]
\end{equation}
It should be noted that, for any value of $\alpha$, $D\tilde{Y}_t$ is
symmetric, i.e. $D \tilde{Y}_t(u_\sigma,u_\sigma') = D
\tilde{Y}_t(u_\sigma',u_\sigma)$ and this is the reflection of the fact that
in this Gaussian case the detailed balance is always satisfied.

\begin{figure}[h]
  \includegraphics[clip=true,height=4cm]{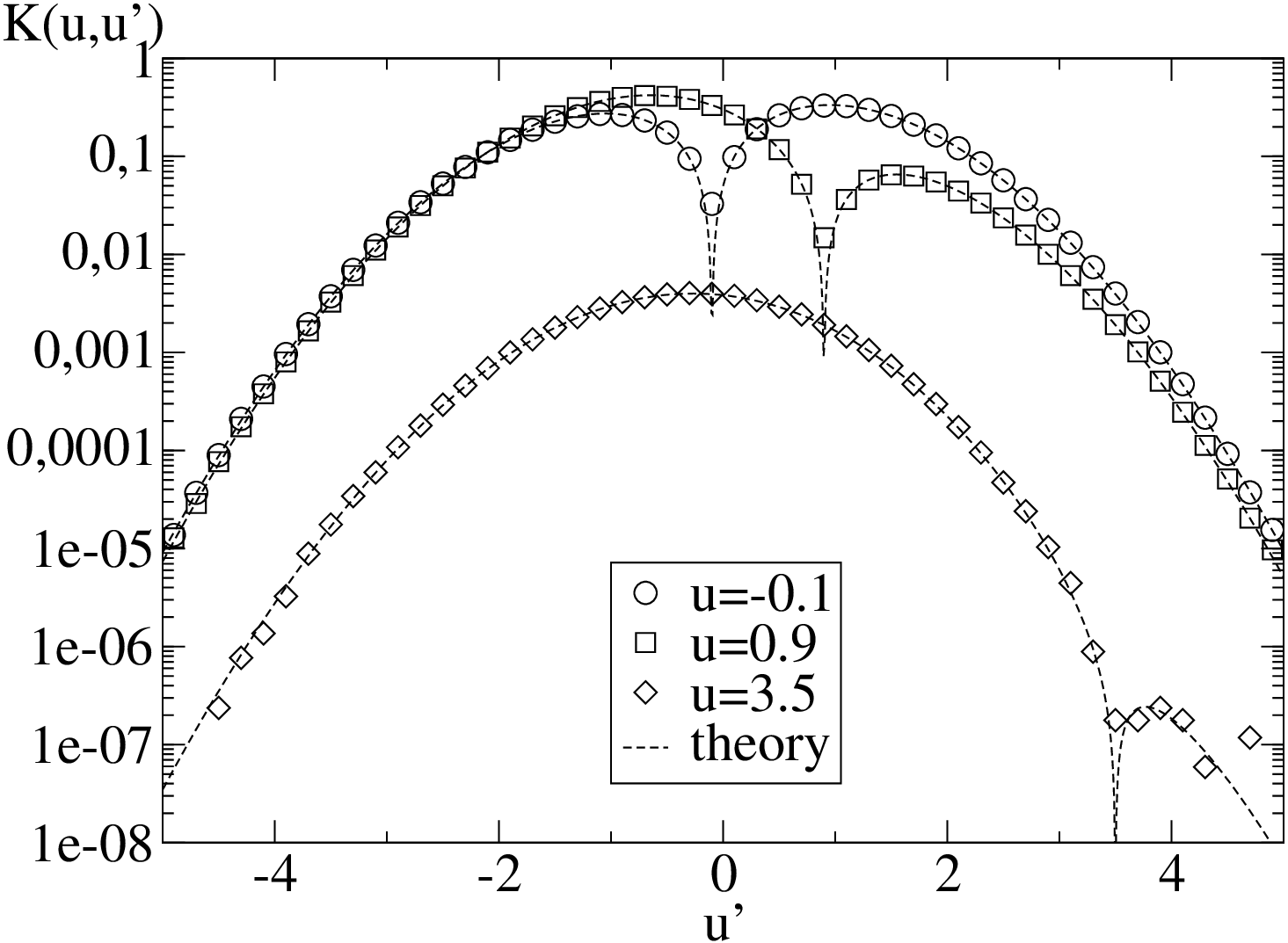}
  \includegraphics[clip=true,height=4cm]{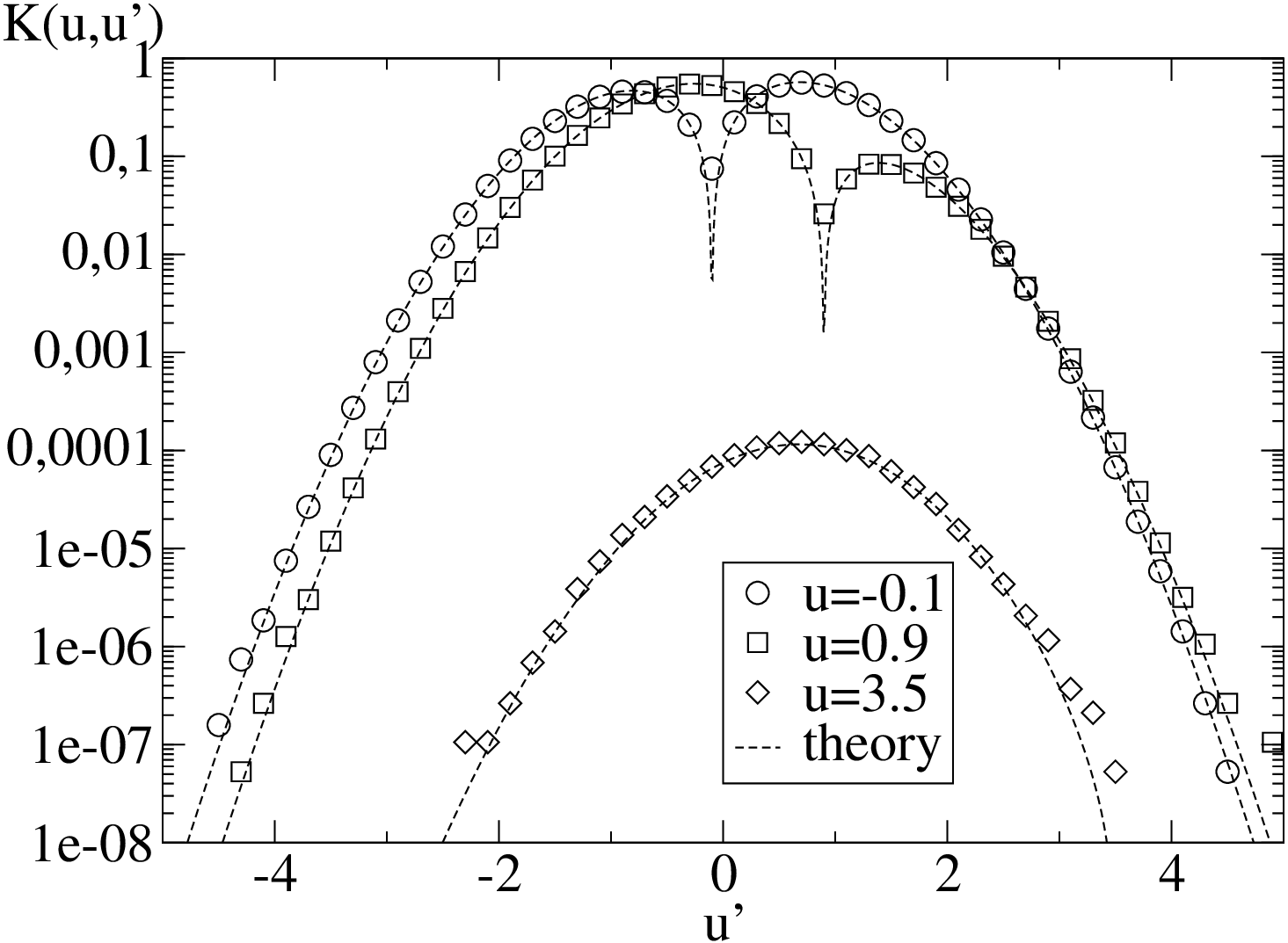}
  \includegraphics[clip=true,height=4cm]{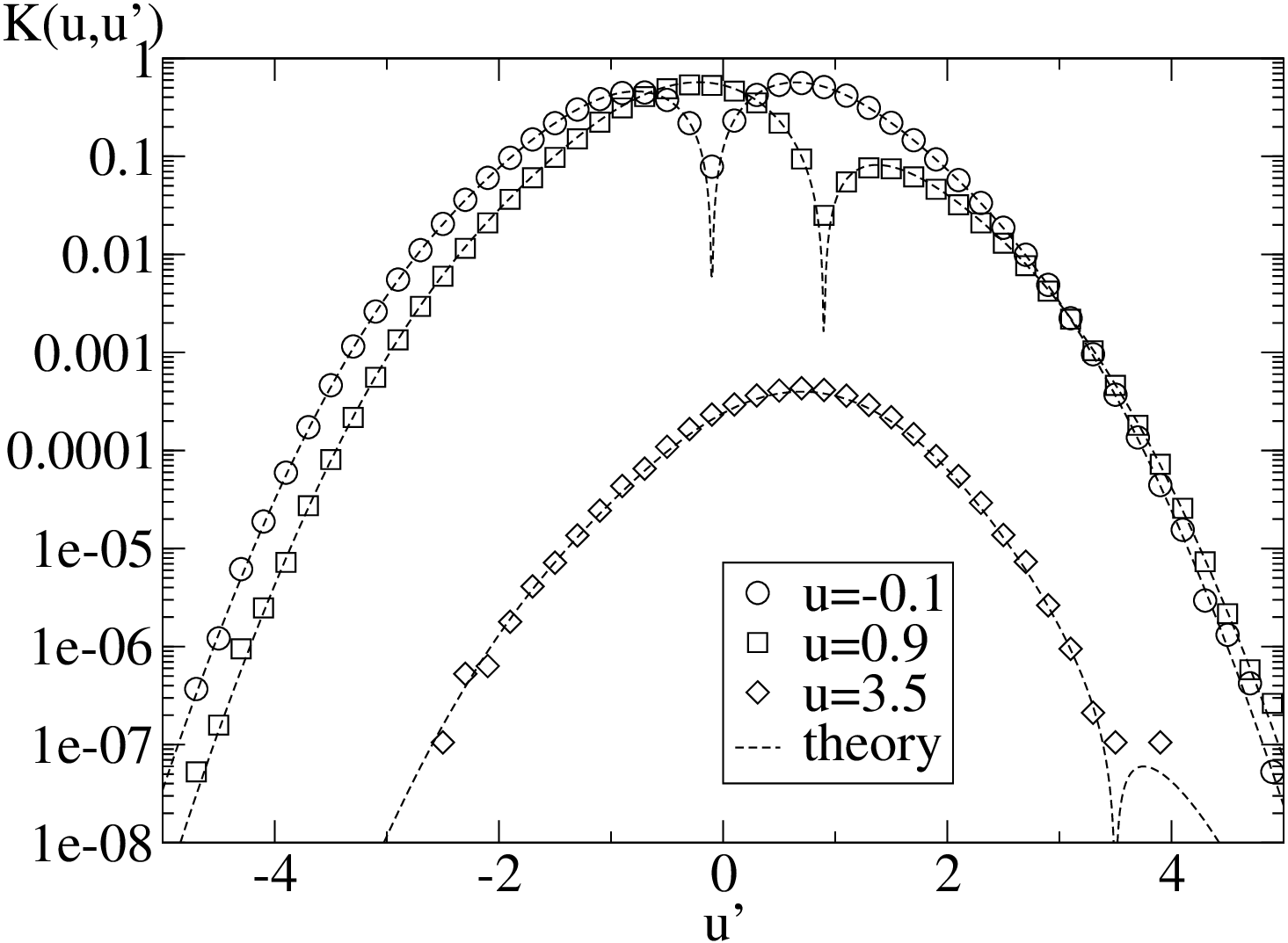}
\caption{
  Measure of transition rates in Monte Carlo simulations. Sections of the
  function $\frac{D\tilde{Y}(u_\sigma,u_\sigma')}{t \dd u_\sigma \dd u_\sigma'}$ for
  the values $u_\sigma=-0.1$, $u_\sigma=0.9$ and $u_\sigma=3.5$ for three
  different cases. Left: the gas has Gaussian velocity pdf with temperature
  $T=1$ and $\alpha=1$ (thermodynamic equilibrium). Center: the gas has a
  Gaussian velocity pdf and $\alpha=0.5$ (detailed balance in absence of
  thermodynamic equilibrium, $T_*=0.6\equiv T'<T$). Right: the gas has
  non-Gaussian velocity pdf characterized by a first Sonine correction with
  $a_2=0.1$ (non-equilibrium stationary state, $T_*=0.62 \approx T'<T$ and
  $a_2^*=0.057 \ll a_2$). The dashed lines mark the theoretical fits given in
  equation~\eqref{eq:gauss1},~\eqref{eq:gauss}
  and~\eqref{eq:nongauss}.\label{fig:gauss1} }
\end{figure}

In left and center frames of figure~\ref{fig:gauss1} the results of DSMC
simulations with Gaussian bulk and $\alpha=1$ and $\alpha=0.5$ are shown for
some sections of the measured $D\tilde{Y}_t$, showing the perfect agreement
with the theoretical fits.

Long calculations lead to a similar result for the non-Gaussian case, when the
gas has a velocity distribution given by a Gaussian, with temperature T,
corrected by the second Sonine polynomial with a coefficient $a_2$. In this
case an exact theory for the stationary state of the tracer particle does not
exist. Nevertheless, numerical simulations show that the velocity distribution
of the tracer, $P_*(v)$ is well approximated by a Gaussian with temperature
$T_*$ corrected by the second Sonine polynomial with a coefficient $a_2^*$. In
particular we have evidence that $T_* \approx T'=\frac{\alpha+1}{3-\alpha}T$
(the Martin-Piasecki temperature, which is exact in the Gaussian case) and
$a_2^* < a_2$. For hard spheres we obtain:

\begin{multline} \label{eq:nongauss}
  \frac{D\tilde{Y}_t(u_\sigma,u_\sigma')}{\dd u_\sigma \dd u_\sigma' t}=
  \left(\frac{2}{1+\alpha}\right)^2\frac{1}{2\sqrt{T T_*}}
  |u_\sigma'-u_\sigma|\exp\left[-\frac{1}{2T}
    \left(\frac{2}{1+\alpha}(u_\sigma'-u_\sigma)
      +u_\sigma\right)^2-\frac{1}{2T_*}u_\sigma^2\right]\\
  \times\left[1+a_2^*\left(\frac{3}{8}-\frac{3}{4}\frac{u_\sigma^2}{T_*}
      +\frac{u_\sigma^4}{8T_*^2}\right)\right]
  \left\{1+a_2S_2^{d=1}\left[\frac{\left(\frac{2}{1+\alpha}
          (u_{\sigma}'-u_{\sigma})+u_{\sigma}\right)^2}{2T}\right]\right\}
\end{multline}

this formula is very well verified by the transition rates observed in
simulations, as shown in the rightmost frame of figure~\ref{fig:gauss1}.

\subsection{Fluctuation Relations}

In this paragraph we want to show the numerical results obtained measuring
$W(\tau)$ and $\overline{W}(\tau)$ in DSMC simulations~\cite{bird}. In
particular our main aim is to verify the Fluctuation Relations $FR_W$ and
$FR_{\overline{W}}$. Actually there are three equivalent relations that can be
written at finite times $\tau$.

\begin{subequations}
\begin{align}
  as_\tau(W)&=\log f^\tau(W)-\log f^\tau(W)=W \label{as_es}\\
  as^*_\tau(w)&=\frac{1}{\tau}[\log f^\tau(\tau w)-\log f^\tau(-\tau w)]=w\\
  as^{**}_\tau(q)&=\frac{1}{\tau \langle w \rangle}[\log f^\tau(\tau \langle w
  \rangle q)-\log f^\tau(-\tau \langle w \rangle q)]=q
\end{align}
\end{subequations}
where $f^\tau(x)$ is the probability density function of finding one of the
two functional after a time $\tau$ equal to $x$ (i.e. $f^\tau_W$ or
$f^\tau_{\overline{W}}$ defined at the beginning of this section), while
$\langle w \rangle$ is a long time average of $W(\tau)/\tau$. The verification
of the GC-LS Fluctuation Relation requires the measure of $as^{**}_\tau(q)$ at
very large values of $\tau$, for values of $q$ at least of order $1$, which
corresponds to measure $as_\tau(W)$ for values of $W$ at least of order $\tau
\langle w \rangle$ at large values of $\tau$. On the contrary the verification
of the ES Fluctuation Relation requires equation~\eqref{as_es} to be true at
all times.  We have chosen to display $as_\tau(W)$ vs. $W$ as well as
$as_\tau(\overline{W})$ vs. $\overline{W}$.

\begin{figure}
  \includegraphics[width=6.2cm,clip=true]{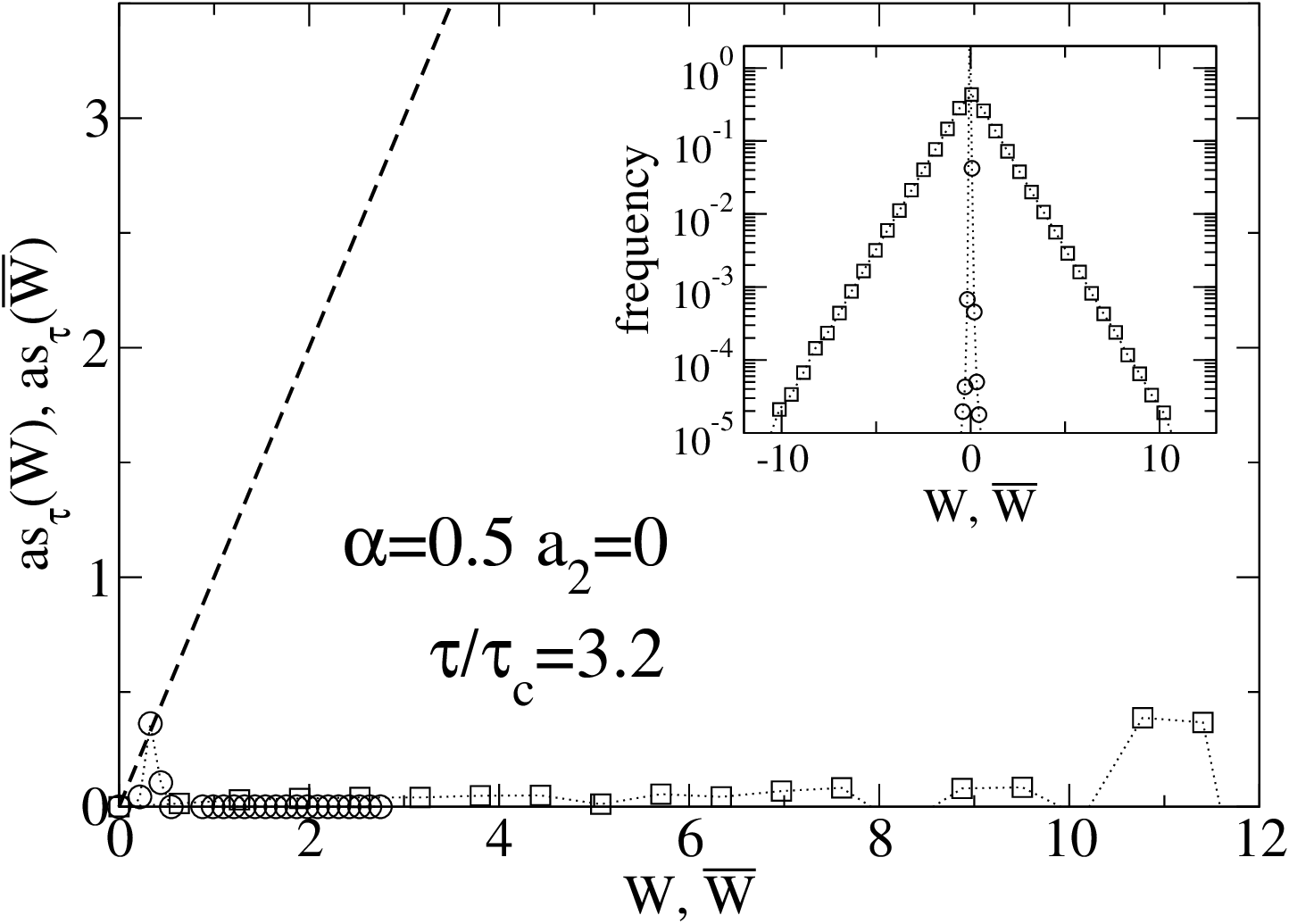}
  \includegraphics[width=6.2cm,clip=true]{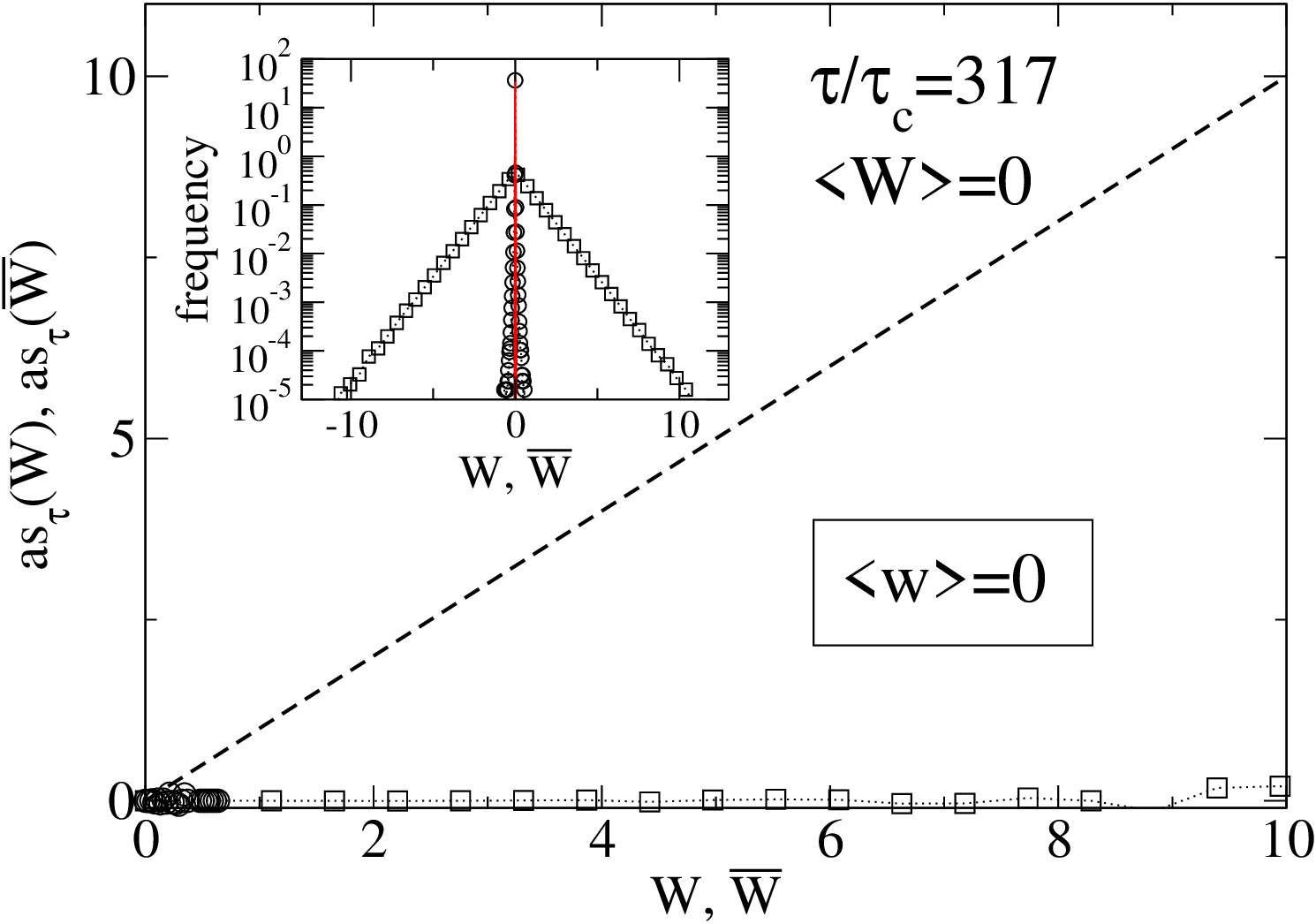}\\
  \includegraphics[width=6.2cm,clip=true]{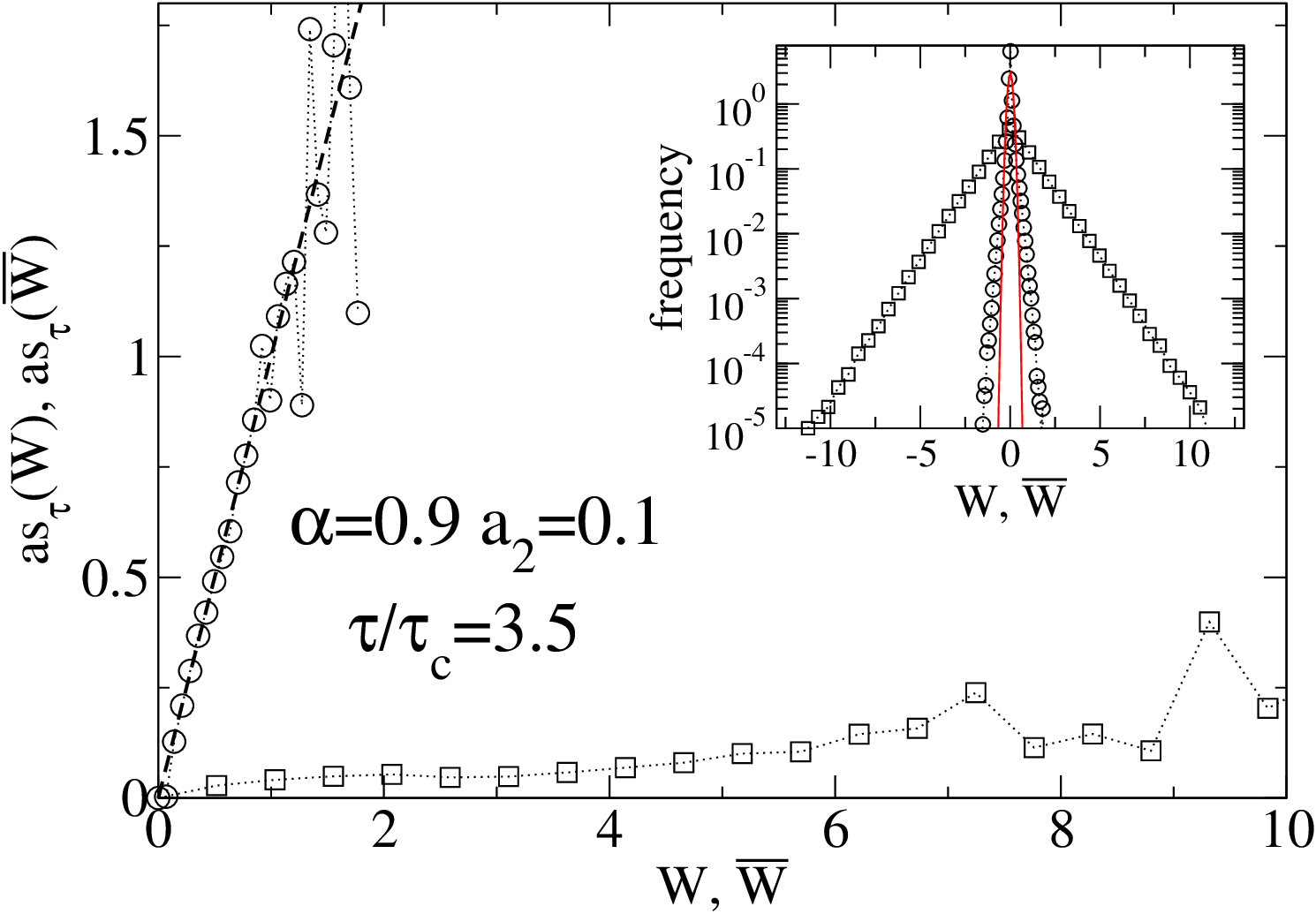}
  \includegraphics[width=6.2cm,clip=true]{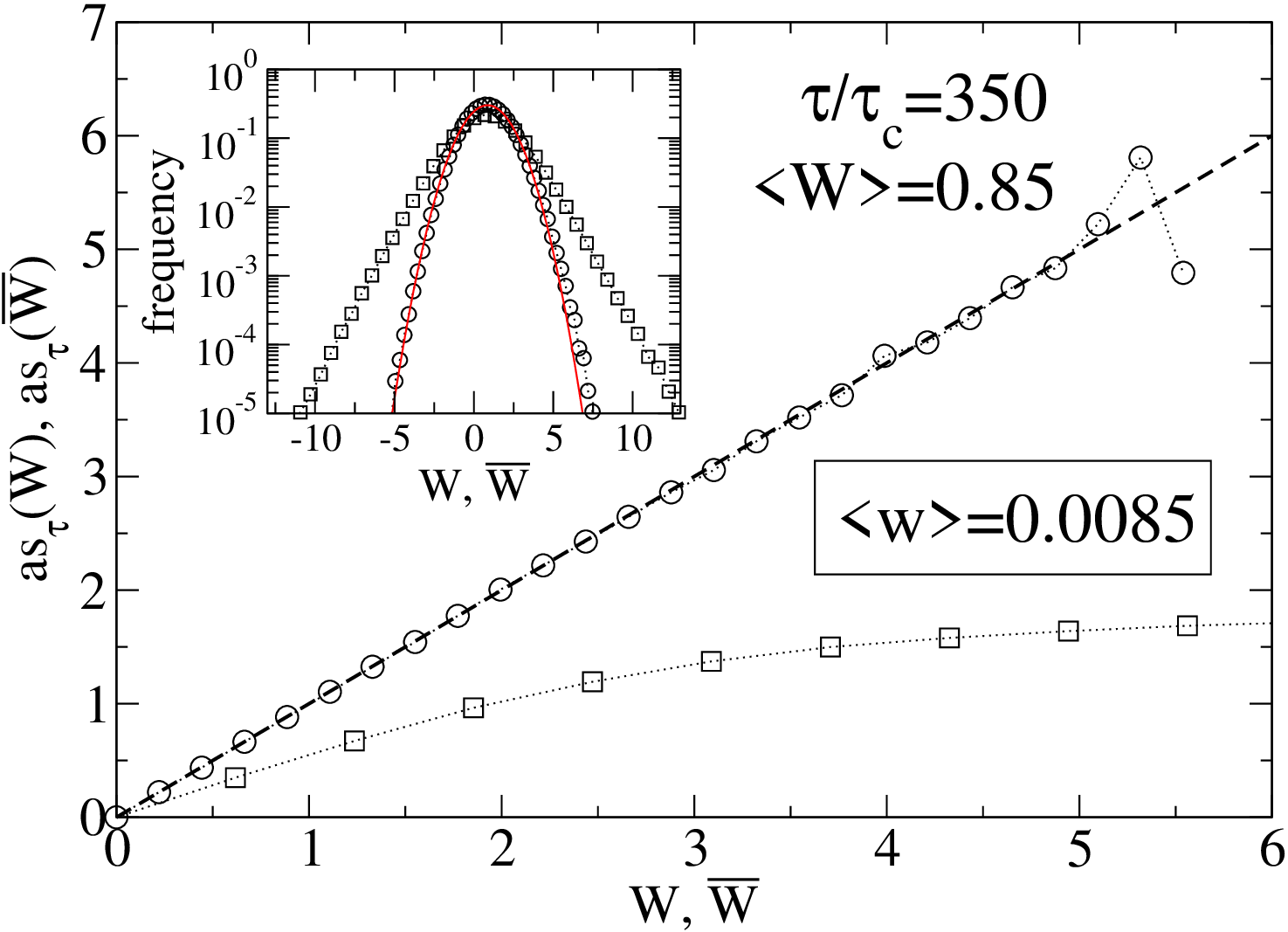}\\
  \includegraphics[width=6.2cm,clip=true]{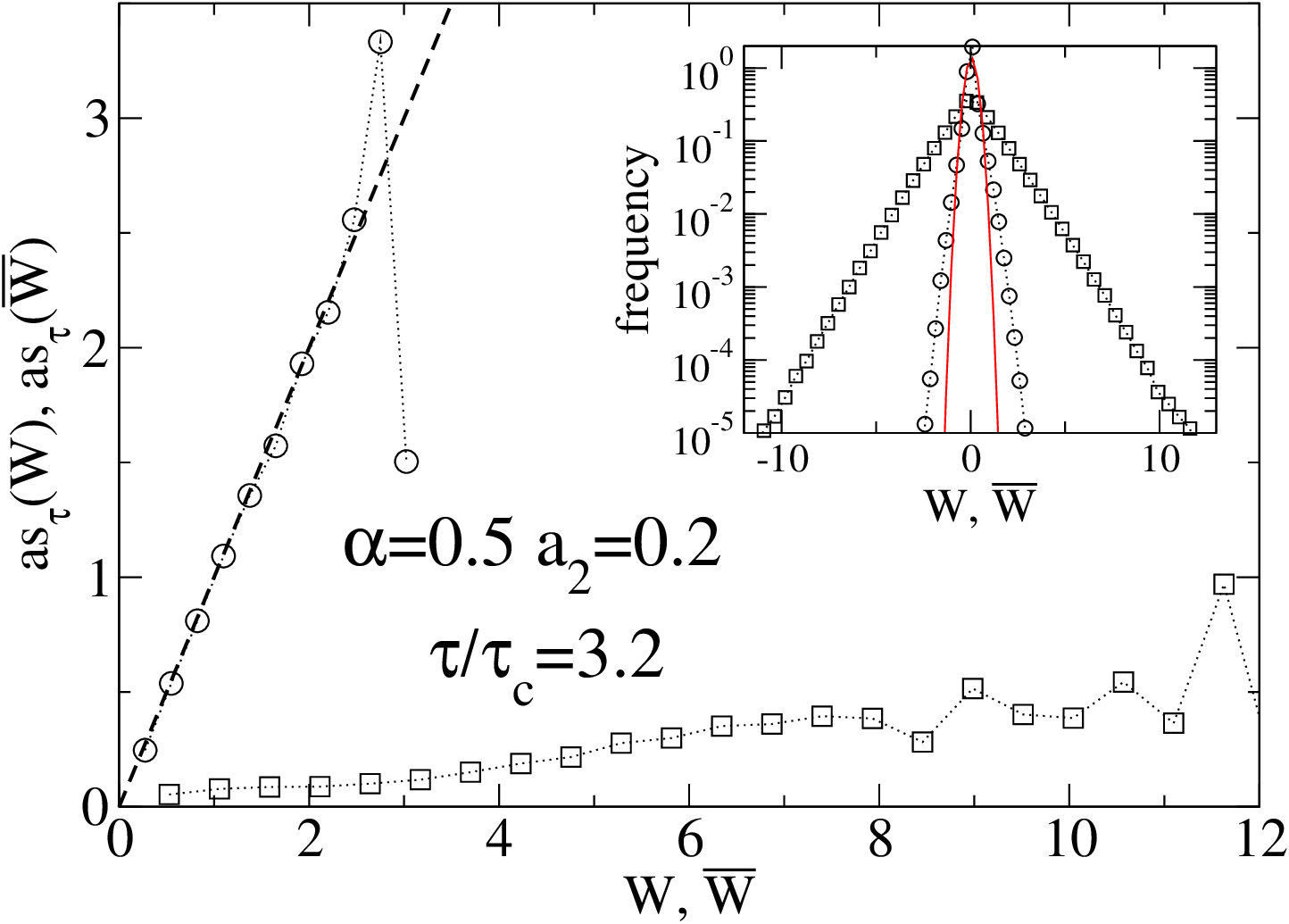}
  \includegraphics[width=6.2cm,clip=true]{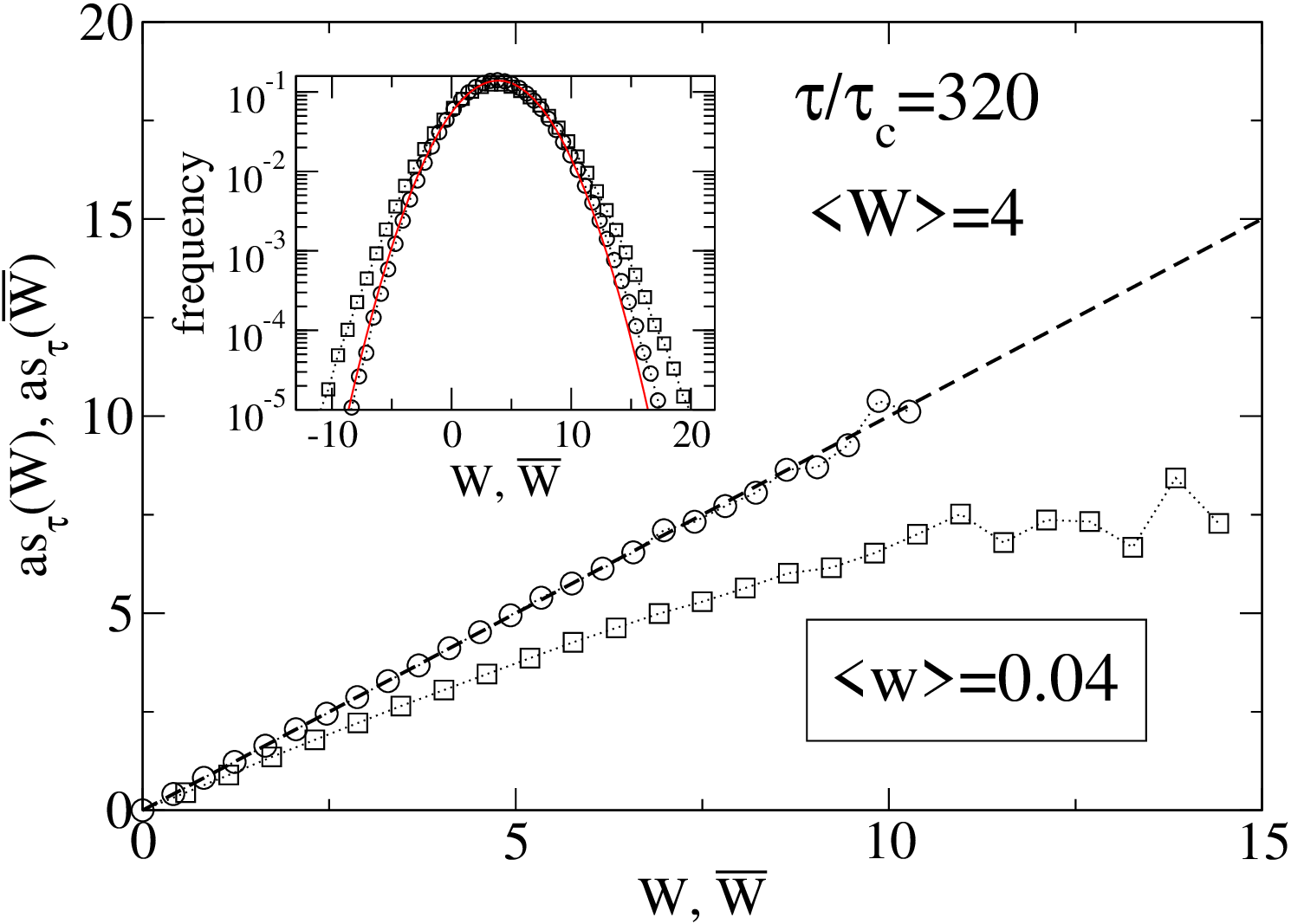}\\
  \includegraphics[width=6.2cm,clip=true]{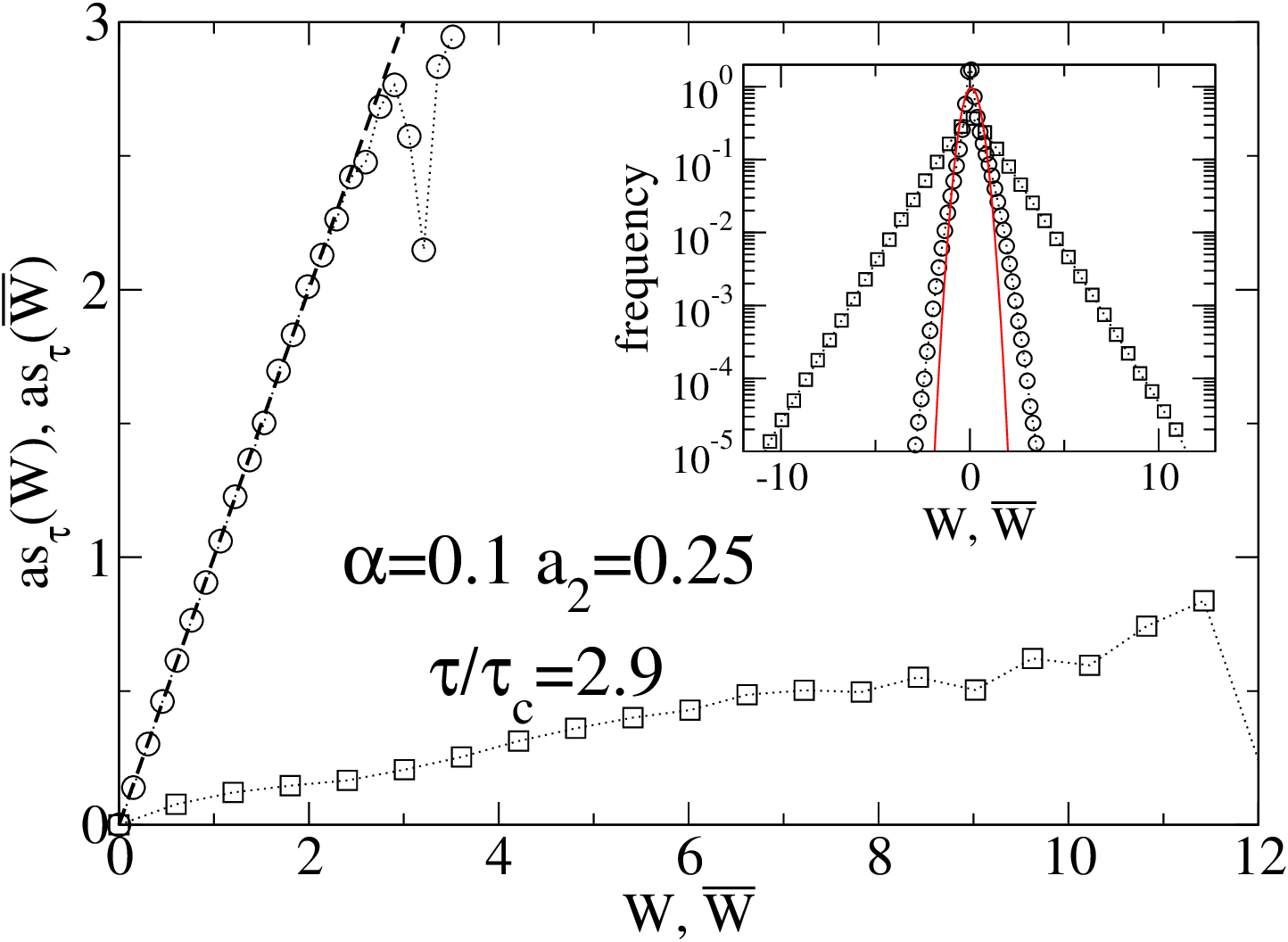}
  \includegraphics[width=6.2cm,clip=true]{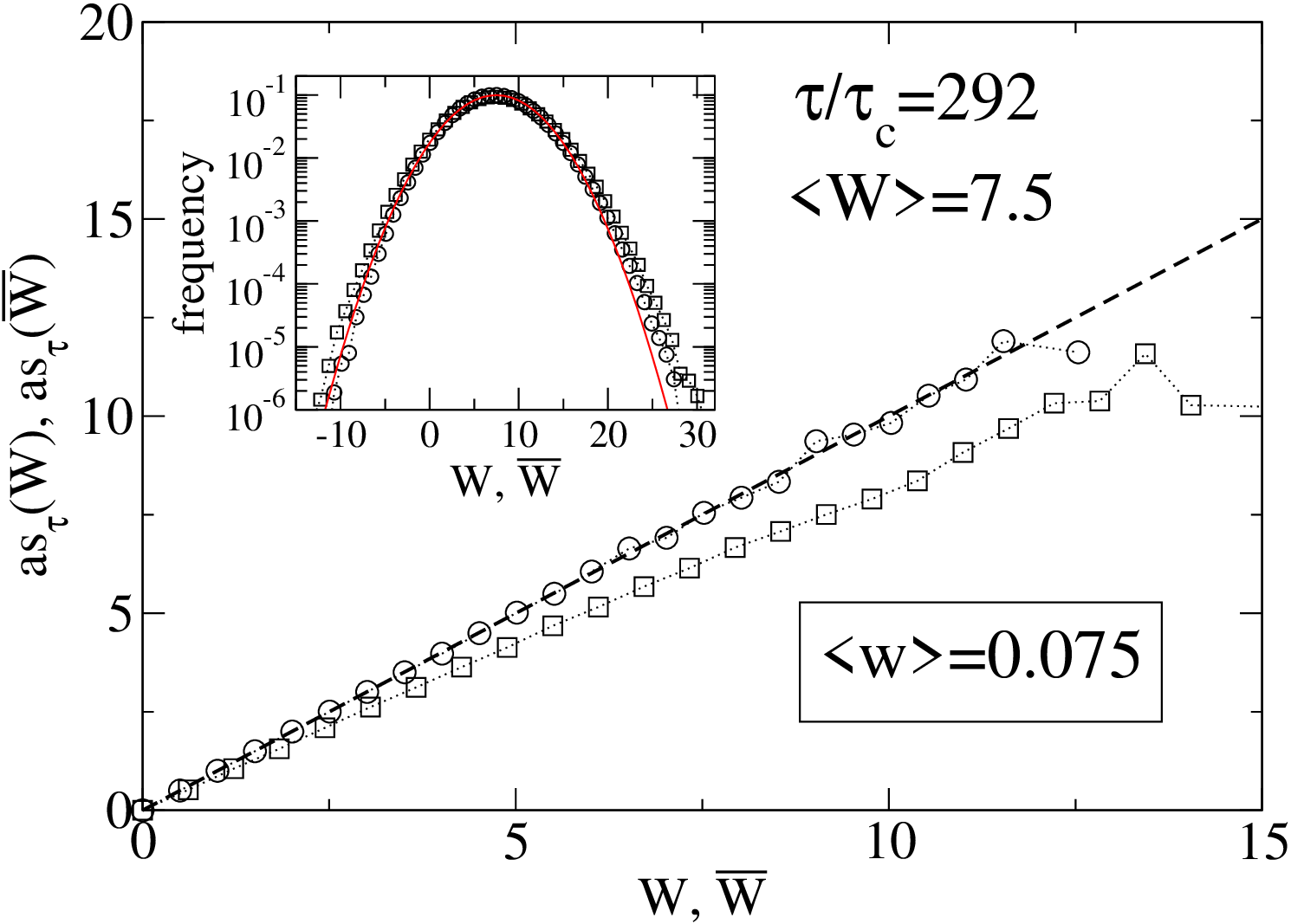}\\
  \includegraphics[width=6.2cm,clip=true]{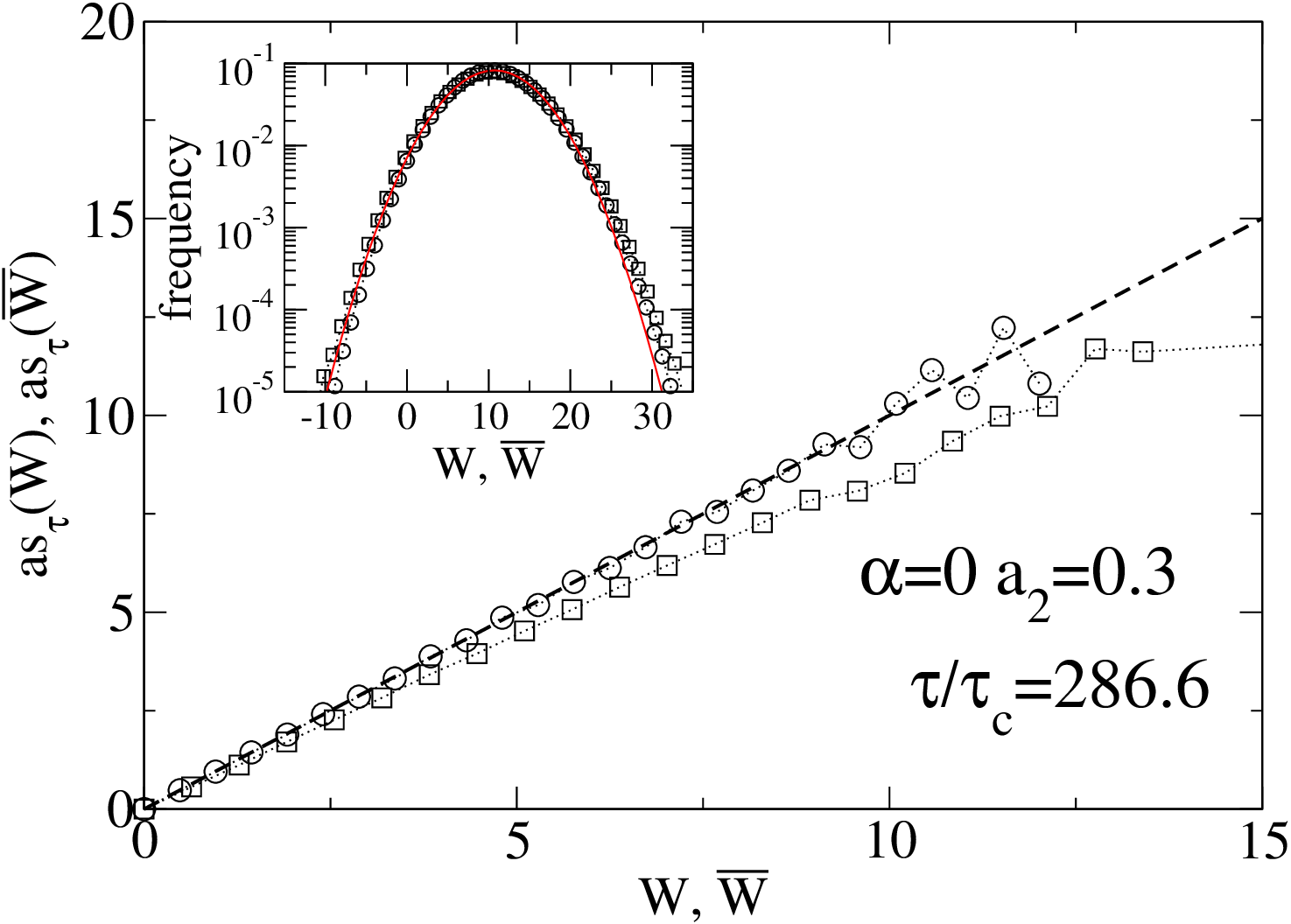}
  \includegraphics[width=6.2cm,clip=true]{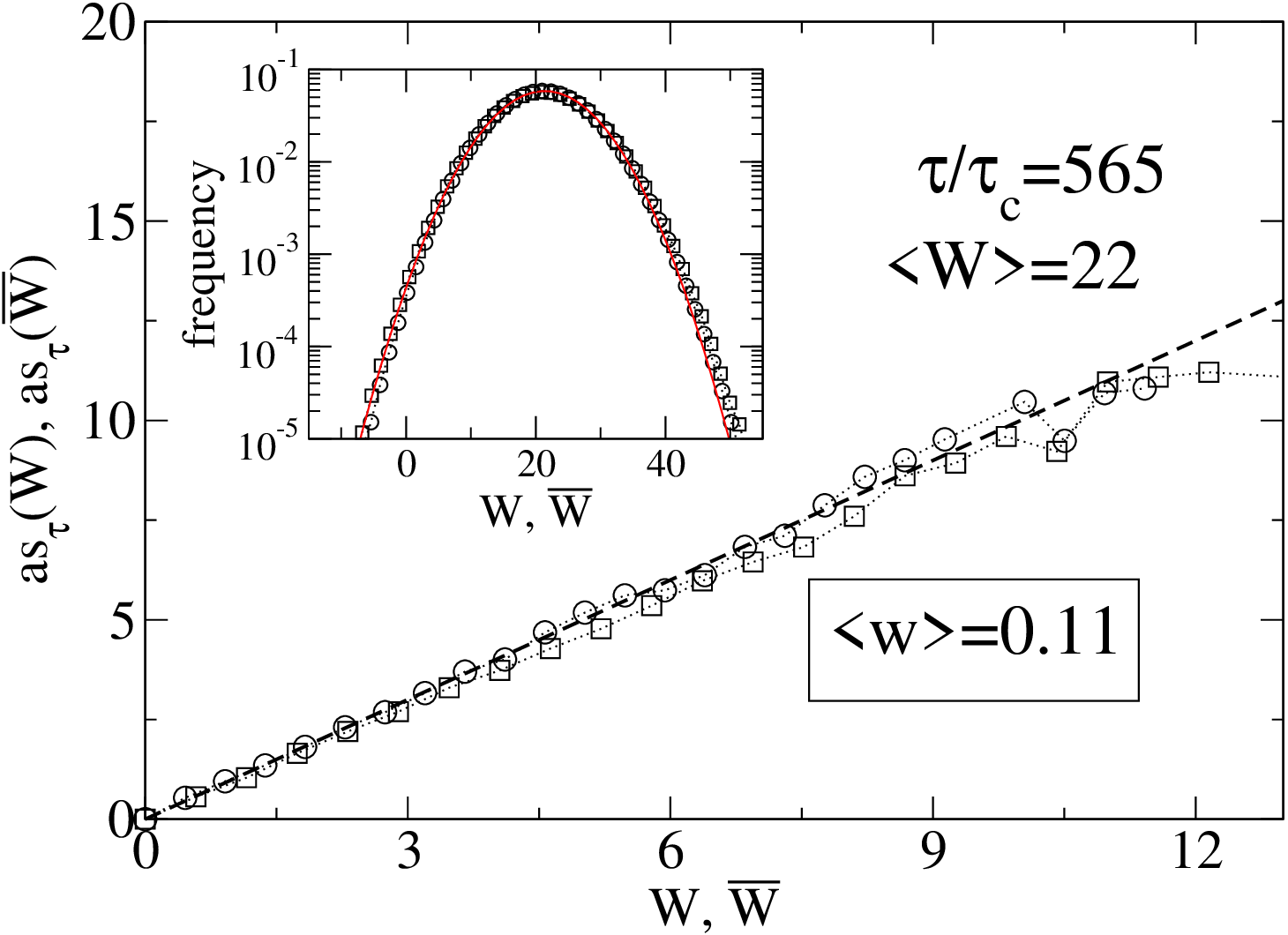}
\caption{Verification of the Fluctuation Relations for $W(\tau)$ (squares) and
  $\overline{W}(\tau)$ (circles). Each line is composed of two graphs and
  shows the results for a particular choice of $\alpha$ and $a_2$ (the
  coefficient of the first Sonine correction characterizing the
  non-Gaussianity of the velocity pdf of the gas): the left graph is at small
  times and the right graph at large times. In each frame the inset contains
  the pdfs of $W$ and $W$, while the main plot shows $as_\tau(W)$ vs. $W$ as
  well as $as_\tau(\overline{W})$ vs. $\overline{W}$. The times $\tau$ are
  rescaled with the tracer mean free time $\tau_c$ (which varies with the
  parameters).  For each choice of parameters, inside the right graph, it is
  shown the value of $\langle W(\tau) \rangle$ and of $\langle w \rangle \equiv \langle W(\tau) \rangle/\tau
  \equiv \langle \overline{W}(\tau) \rangle/\tau \equiv \lim_{\tau \to \infty}
  W(\tau)/\tau \equiv \lim_{\tau \to \infty} \overline{W}(\tau)/\tau$ which
  marks the distance from equilibrium. The dashed line has slope
  $1$.\label{fig:pdf_gc}}.
\end{figure}

To measure the quantity $\overline{W}(\tau)$ on each segment of trajectory it
is assumed that $P_*(\mathbf{V})$ is a Gaussian with a first Sonine correction
and the values $T_*$ and $a_2^*$ are measured during the simulation itself and
used to compute the ``boundary term'' appearing in the definition of
$\overline{W}$, eq.~\eqref{entropyb}. All the results are shown in
figure~\ref{fig:pdf_gc}, for many different choices of the time $\tau$ and of
the parameters $\alpha$ and $a_2$. We display in each figure the value of
$\langle w \rangle=\lim_{\tau \to \infty} W(\tau)/\tau$. Following the idea
that this number is an average entropy production rate, we use it to measure
how far from equilibrium our system is. This idea is well supported by the
results of our simulations: $\langle w \rangle$ is zero when $a_2=0$ and
increases as $\alpha$ is decreased and $a_2$ is increased. We recall that
$\lim_{\tau \to \infty} W(\tau)/\tau=\lim_{\tau \to \infty}
\overline{W}(\tau)/\tau\equiv \langle w \rangle$, since the difference between
the two functionals has zero average at large times $\tau$. We also remark
that the distribution of both quantities $W$ and $\overline{W}$ are symmetric
at equilibrium (i.e. when $a_2=0$).

We first discuss the results concerning the fluctuations of $W(\tau)$,
identified in figure~\ref{fig:pdf_gc} by squared symbols. The pdf of
these fluctuations are shown in the insets. They are strongly
non-Gaussian, with almost exponential tails, at low values of $\tau$
for any choice of the parameters.  At large values of $\tau$ (many
hundreds mean free times), the situation changes with how far from
equilibrium the system is. At low values of $\langle w \rangle$ the
tails of the distribution are very similar to the ones observed at
small times. At higher values of $\langle w \rangle$ the distribution
changes with time and tends to become more and more Gaussian. We must
put a great care in the interpretation of this observation. There is
not a firm guiding principle to determine how large the time $\tau$
and the values of $W$ should be, in order to claim that ``large
deviations'' are being probed. A simplistic approach is the following:
every time one observes non-Gaussian tails then one concludes that
these tails are, in fact, large deviations, since ``small deviations''
should be Gaussian. Anyway this approach is misleading: a sum of
random variables can give non-Gaussian tails if the number of summed
variables is low and the variables itself have a non-Gaussian
distribution. Therefore time plays a fundamental role and cannot be
disregarded. The observation of non-Gaussian tails alone is
useless. We will discuss in the next section the problems related to
the true asymptotics of the pdf of $W(\tau)$. Here we stick to the
mere numerical observations: we have not been able to measure negative
deviations larger than the ones shown in
figure~\ref{fig:pdf_gc}. Distributions obtained at higher values of
$\tau$ yielded a smaller statistics and very few negative
events. Using the distributions at hand, we could not verify the GC-LS
Fluctuation Relation.  In the last case, which is very far from
equilibrium ($a_2=0.3$, $\alpha=0$), a behavior compatible with the
GC-LS Fluctuation Relation is observed, i.e. $as_\tau(W) \approx W$,
but the range of available values of $W$ is much smaller than $\langle
W \rangle$. At this stage, and in practice, we consider such results a
failure of the GC-LS Fluctuation Relation for continuous Markov
processes.

On the other hand, things for the functional $\overline{W}(\tau)$ are much
easier. The pdfs of $\overline{W}(\tau)$ always yield deviations from the
Gaussian, but weaker than those observed for $W(\tau)$. The ES Fluctuation
Relation is always fairly satisfied in all non-equilibrium cases at all times
$\tau$. The verification of both Fluctuation Relations is meaningless in the
equilibrium cases ($a_2=0$) because $as_\tau \equiv 0$.

\subsection{Difference between the fluctuations of $W(\tau)$ and of $\overline{W}(\tau)$}

As already noted, the two action functionals defined in
formula~\eqref{entropy} differ for the presence/absence of a {\em boundary
  term}. In principle one should expect that it is not possible to distinguish
between the large deviations (i.e. the leading factor of the pdf at large
times) of $W(\tau)$ and $\overline{W}(\tau)$. Anyway there are no arguments to
predict the threshold time above which the two functionals (i.e.  most of
their fluctuations) become indistinguishable from the point of view of large
deviations~\cite{rondoni,galla_last}. Moreover, to our knowledge, it is
not proved that the large deviation functions of $W$ and $\overline{W}$
coincide. In this paragraph we first discuss numerical observation: they
concern, generically, ``deviations'', i.e. fluctuations at finite time and
finite values of the variables. Finally we also argue about the asymptotic
behavior of the pdfs of $W$ and $\overline{W}$. As already stated, in
numerical experiments, averages have been always obtained using many
independent segments of length $\tau$ from a very long trajectory. This
amounts to sampling the stationary velocity pdf of the tracer.

\begin{figure}[htbp]
\includegraphics[width=12cm,height=10cm,clip=true]{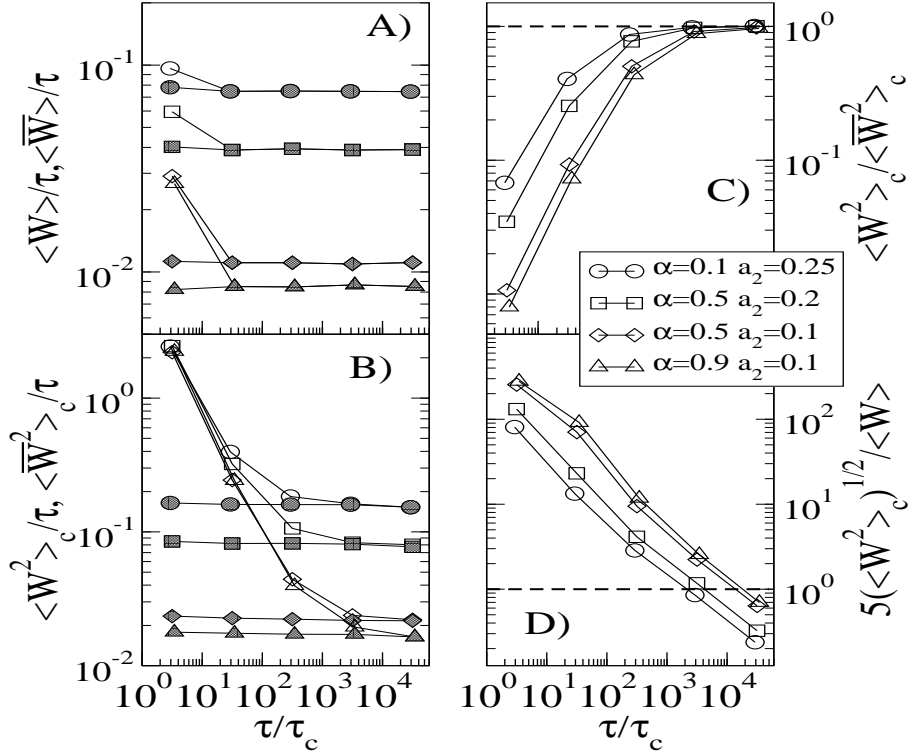}
\caption{Cumulants of the fluctuations of $W(\tau)$ (empty symbols) and
  $\overline{W}(\tau)$ (full symbols). A): average values of $W(\tau)$ and
  $\overline{W}(\tau)$ as a function of $\tau$ and for different choices of
  the parameters $\alpha$ and $a_2$. B) second cumulants of $W(\tau)$ and
  $\overline{W}(\tau)$ as a function of $\tau$. C): ratio between the second
  cumulant of $\overline{W}(\tau)$ and that of $W(\tau)$ as a function of
  $\tau$. D): ratio between $5\sqrt{\langle W(\tau)^2 \rangle_c}$ and $\langle
  W(\tau) \rangle$. When this ratio is much smaller than $1$ the probability
  of observing negative events in the fluctuations of $W(\tau)$ becomes
  extremely low. All the times $\tau$ are rescaled by the tracer mean free
  time.}\label{fig:stat}
\end{figure}

In figure~\ref{fig:stat} we have summed up the results of numerical
simulations with different values of $\alpha$, $a_2$ and $\tau$. The averages
$\langle W(\tau)\rangle$ and $\langle \overline{W}(\tau)\rangle$ converge to
the same value in a time smaller than $100$ average collision times (frame 'A'
of figure~\ref{fig:stat}). The analysis of fluctuations (frames 'B' and 'C'), by
means of the measure of their variance, instead indicates that the convergence
is much more slower. Remarkably the convergence is slower when the system is
closer to equilibrium.  We also tried to answer to the question: is it
possible to verify the Fluctuation Relation $FR_W$ at finite times in the
regime (large $\tau$) where the two functionals have negligible differences?
The main obstacle to the verification of a Fluctuation Relation is the lack of
negative events at large values of $\tau$. The probability of finding a
negative event is equivalent to the probability of finding a negative
deviation from the mean of the order of the mean itself. We have seen that
when the pdf of $W(\tau)$ seems to converge to its asymptotic scaling form,
its tails are not far from those of a Gaussian and the bulk (i.e. its ``small
deviation'' range) is clearly Gaussian. This means that there is a very
sensitive decay of probability in the range of a finite number of standard
deviations from the mean. From a numerical point of view it is very unlikely
to observe values of $W(\tau)$ below the value $\langle W(\tau)
\rangle-5\sqrt{\langle W^2(\tau) \rangle_c}$, where $\langle
W^m(\tau)\rangle_c$ is the $m$-th cumulant of $W(\tau)$, or, equivalently,
that it is very unlikely to observe negative events if $5\sqrt{\langle
  W^2(\tau)\rangle_c} \ll \langle W(\tau) \rangle$. Frame 'D' of
figure~\ref{fig:stat} shows the ratio between $5\sqrt{\langle W^2(\tau)
  \rangle_c}$ and $\langle W(\tau) \rangle$. The conclusion of this analysis
is that when $\tau$ is large enough to guarantee, at least at the level of the
second cumulant, the convergence of the pdf of $W$ (or its convergence to the
pdf of $\overline{W}$), the probability of observing negative events is so
small that a huge statistics is required, in order to make reliable
measurement of the Fluctuation Relation.

We now discuss one possible reason for such a slow convergence of the pdf of
the fluctuations of $W(\tau)$. We will call it the ``boundary term
catastrophe''. As a matter of fact it seems that $W(\tau)$, even at very large
values of $\tau$, more evidently in the close-to-equilibrium cases, {\em
  remembers} its own small time fluctuations. This is clearly seen in
figure~\ref{fig:pdf_gc}, four topmost frames. The tails of $W(\tau)$ at large
$\tau$ are almost identical to those at small $\tau$. What is happening? If
one looks closely at expression~\eqref{ratio_sonine}, which is summed over
many collisions to give $W(t)$, discovers that
\begin{equation} \label{sum}
W(t)=\frac{3-\alpha}{1+\alpha}\frac{|\mathbf{v}_1|^2
-|\mathbf{v}_{n(t)}|^2}{2T}+\sum \log \frac{1+a_2 ...}{1+a_2 ...}
\end{equation}
where we have used a shortened notation to identify the ``non-equilibrium''
part which is the sum over all the collisions in the time $\tau$ of the
logarithms of ratios between the Sonine contributions. The sum of the energy
differences reduces to a difference between the first and the last term, which
we can call again a ``boundary term''. It easy to realize that the pdf of this
term alone, being the pdf of a difference of energies, has exponential tails
(in granular gases they can be even slower). This is the dominant term in the
fluctuations of $W(\tau)$ at small $\tau$ and it can dominate the fluctuations
of $W(\tau)$ even at very large $\tau$, depending on the amplitude of the
fluctuations of the ``Sonine'' term. The ``Sonine'' term is small near
equilibrium and increases as the system gets far from equilibrium. This
explains why the memory is stronger (i.e. convergence is lower) as the system
is closer to equilibrium.

The situation dramatically changes when looking at the second functional,
$\overline{W}(\tau)$. The ``boundary term'' discussed above almost perfectly
annihilates with the boundary term appearing in the definition of
$\overline{W}(t)$. Following the numerical observation that the pdf $P_*$ of
the velocity of the tracer is almost Gaussian with temperature $T_* \approx
T'\equiv \frac{1+\alpha}{3-\alpha}T$, in fact we see that

\begin{equation}
\log \frac{P_*(\mathbf{v}_1)}{P_*(\mathbf{v}_{n(t)})} 
\approx \frac{-|\mathbf{v}_1|^2+|\mathbf{v}_{n(t)}|^2}{2T_*} 
\approx \frac{3-\alpha}{1+\alpha}
\frac{-|\mathbf{v}_1|^2+|\mathbf{v}_{n(t)}|^2}{2T}
\end{equation}
The error in this approximation is of the same order of one of the Sonine
terms appearing in the sum~\eqref{sum} and is therefore subleading at large
times. This crucial observation explains the absence of exponential tails in
the pdf of $\overline{W}(t)$.

It should be stressed that this ``boundary term catastrophe'' has relevant
consequences not only in the realm of numerical simulations, but also from the
point of view of large deviation theory. We suspect that, even with a computer
of infinite power, i.e. with the ability of measuring with infinite quality
the pdf of $W(\tau)$ at any time $\tau$, the results of a large deviation
analysis would yield similar results: the exponential tails related to the pdf
of the $v_1^2-v_n^2$ are never forgotten. This is the consequence of the
following simple observation: following the definition of ``large deviations
rate'' of a pdf $f_t$ which depends upon a parameter $t$, we can also
calculate the large deviations rate of a pdf $F$ which {\em does not depend}
upon that parameter. We recall the definition:

\begin{equation}
\pi(w)=\lim_{t\to \infty}\frac{1}{t}\log{f_t(t w)}.
\end{equation}
It can be immediately seen that such definition gives a {\em finite} result if
applied to a pdf $F$ with exponential tails, even if it does not depend upon
$t$. This is equivalent to state that exponential tails always contribute to
large deviations. The apparent loss of memory that can be observed in the
strong non-equilibrium cases of figure~\ref{fig:pdf_gc} is only due to the
limited range available in the measure of the pdf: these tails, in fact, get
farther and farther as time increases. Moreover we recall that such
exponential tails are associated to the distribution of energy fluctuations
and are always present in a system with a finite number of particles, not only
in the $1$-particle case. They disappear only in the thermodynamic limit or in
the presence of physical boundaries.  We conclude this discussion mentioning
that the boundary term catastrophe is responsible for the failure of GC
Fluctuation Relation in other systems~\cite{rondoni,farago,vanzon,galla_last}.

\section{Conclusions}

In conclusion we have studied the dynamics of a tracer particle in a driven
granular gas. The tracer performs a sequence of collisions changing its
velocity, thus performing a Markov process with transition rates that can be
obtained from the linear Boltzmann equation. We have given a general
expression for these rates, and we have also calculated them for a number of
possible interesting cases. There is a major discrimination between the
Gaussian and the non-Gaussian case, depending upon the velocity distribution
of the gas particles. In the Gaussian case the tracer is at equilibrium, even
if the collisions are dissipative and energy equipartition is not satisfied.
In the non-Gaussian case the tracer reaches a statistically stationary
non-equilibrium state: in this state a dissipative flow can be measured in the
form of an action functional that has been defined by Lebowitz and
Spohn~\cite{lebowitz}. The average of this flow well describes the distance
from equilibrium. Moreover the rate of variation of the entropy of the
tracer can be split up in two contributions, one always positive and the
second given by the average of this dissipative flux. Following the Lebowitz
and Spohn ``recipe'' we have tried to verify the Gallavotti-Cohen fluctuation
relation for the action functional, realizing that this verification is not
accessible. At the same time we have perfectly verified a second fluctuation
relation, for a slightly different functional, which we have called
Evans-Searles Fluctuation Relation because of a formal analogy with it.  We
have discussed the possibility that a non-extensive term (in time) in the
definition of the Lebowitz-Spohn functional can spoil the applicability of the
large deviation principle. In our numerical simulations we have also very well
verified the correctness of the formula for the transition rates, showing a
way to optimize their measure in real experiments. A key point of our work is
in fact that such Lagrangian approach in granular gas theory should be applied
to experiments on dilute granular matter, in order to test directly the limits
of the different Fluctuation Relations. Future work will be directed to the
study of transients and non-homogeneous situations.  The Evans-Searles
Fluctuation Relation is expected to hold also in non stationary regimes, such
as the transient relaxation of a packet of non-interacting tracer granular
particles prepared in some non-typical initial state. Non-homogeneous
situations, such as a boundary driven granular gas, are characterized by the
presence of spatially directed fluxes. It will be interesting to study the
``action functional'' defined in this work and compare it with other measures
of transport.

\noindent {\it Acknowledgments.--} The authors acknowledge A. Barrat
for several fruitful discussions. A. P. warmly thanks L. Rondoni for
discussions and a careful reading of the paper, and acknowledges the
Marie Curie grant No.  MEIF-CT-2003-500944.

\section*{Appendix. Derivation from the linear Boltzmann Equation.}


We first discuss in detail what happens in a collision and then give a
rigorous derivation of the master equation. The collision rule for inelastic
hard spheres reads:

\begin{equation}
\mathbf{v}_1'=\mathbf{v}_1-\frac{1+\alpha}{2} (\bv_{12} \cdot \bs) \bs
\end{equation}
where $\bs$ is the direction
joining the centers of the two colliding particles. There are some
consequences of the collision rules which have to be remarked. 
For simplicity we assume to be in dimension $d=2$.

\begin{itemize}

\item
$\Delta \mathbf{v}_1=\mathbf{v}_1'-\mathbf{v}_1$ is parallel to
$\hat{\mathbf{\sigma}}$, i.e. $\theta=\arctan \frac{\Delta
v_{1y}}{\Delta v_{1x}}$ where $\hat{\sigma}_x=\cos \theta$ and
$\hat{\sigma}_y=\sin \theta$. This is equivalent to recognize that the
velocity $\mathbf{v}_1$ changes only in the direction
$\hat{\sigma}$. The fact that $\mathbf{v}_{12\sigma}$ must be negative
determines completely the angle $\theta$, i.e. the unitary vector
$\hat{\mathbf{\sigma}}$. From here on, we call $\Delta v_1\equiv\Delta
v_{1\sigma}\equiv\Delta \mathbf{v}_1 \cdot \hat{\mathbf{\sigma}}$.

\item
$v_{2\sigma} \equiv \mathbf{v}_2 \cdot \hat{\mathbf{\sigma}} = \frac{2}{1+\alpha}\Delta v_1+v_{1\sigma}=\frac{2}{1+\alpha}v_{1\sigma}'-\frac{1-\alpha}{1+\alpha}v_{1\sigma}$. 

\item 
From the previous two remarks, it is clear that $\Delta v_1$
determines univocally $\hat{\sigma}$ and $v_{2\sigma}$. The component
of $\mathbf{v}_2$ which is not determined by $\Delta v_1$ is the one
orthogonal to $\hat{\mathbf{\sigma}}$. We call $\hat{\mathbf{\tau}}$
the direction perpendicular to $\hat{\mathbf{\sigma}}$, i.e. the
vector of component $(-\sin \theta, \cos \theta)$.  We define
$v_{2\tau}=\mathbf{v}_2 \cdot \hat{\mathbf{\tau}}$. 

\end{itemize}

From the above discussion, it is easy to understand that the
transition probability for the particle $1$ to change velocity during
a collision, going from $\mathbf{v}_1$ to $\mathbf{v}_1'$ must be

\begin{subequations} 
\begin{align}
K(\mathbf{v}_1 \to \mathbf{v}_1') &= C(\mathbf{v}_1,\mathbf{v}_1') \int \dd v_{2\tau} P(\mathbf{v}_2)\\
\mathbf{v}_2&=v_{2\sigma}\hat{\mathbf{\sigma}}+v_{2\tau}\hat{\mathbf{\tau}}\\
\hat{\mathbf{\sigma}}&=(\cos \theta, \sin \theta)\\
\hat{\mathbf{\tau}}&=(-\sin \theta, \cos \theta)\\
\theta&=\arctan \frac{\Delta v_{1y}}{\Delta v_{1x}}\\
\hat{\mathbf{\sigma}} &\parallel \mathbf{v}_1'-\mathbf{v}_1\\
v_{2\sigma}&= \frac{2}{1+\alpha}\Delta v_1+v_{1\sigma}
\end{align}
\end{subequations}
where $P(\mathbf{v})$ is the 1-particle probability density function
for the velocity in the bulk gas. The constant of proportionality $C$
must be of dimensions $1/length$ so that $K$ has dimensions
$1/(velocity^d time)$ which is expected because $K$ is a rate of
change of the velocity pdf (in $d$ dimensions).

Now, we want to obtain the complete result, that is rigorously transform the usual
linear $\gamma$-Boltzmann equation for inelastic models in a Master Equation for a
single-particle Markov chain, i.e. as

\begin{equation}
\frac{\dd P_*(\mathbf{v},t)}{\dd t}=\int \dd \mathbf{v}_1 P_*(\mathbf{v}_1) K(\mathbf{v}_1 \to \mathbf{v}) - 
\int \dd \mathbf{v}_1 P_*(\mathbf{v}) K(\mathbf{v} \to \mathbf{v}_1).
\end{equation}
with $P_*(v)$ the velocity pdf of the test particle.


From this definition (and from the usual linear Boltzmann equation for models
containing a term $|\mathbf{v}_{12}\cdot
\bo|^{\gamma}$, equation~\eqref{boltzmann}) it
follows that

\begin{equation}
K(\mathbf{v}_1 \to \mathbf{v}_1')=\frac{v_0^{1-\gamma}}{l_0}\int \dd \mathbf{v}_2\int \dd \bo \Theta(\bv_{12} \cdot \bo) |\bv_{12} \cdot \bo|^{\gamma} P(\mathbf{v}_2)\delta\left\{\mathbf{v}_1'-\mathbf{v}_1+\frac{1+\alpha}{2} [\bv_{12} \cdot \bo]\bo\right\}.
\end{equation}
where $P(v)$ is the velocity pdf of the bulk gas.
Using that for a generic $d$-dimensional vector $\br=r \brh$ one has 
$\delta (\br - \br_0)= \frac{1}{r_0^{d-1}} \delta (r-r_0) \delta (\brh - \brh_0)$, the previous expression can 
be rewritten as:
\begin{equation}
K(\bv_1 \to \bv_1')= \frac{v_0^{1- \gamma}}{l_0} \int \dd \bv_2 \int \dd \bo \Theta(\bv_{12} \cdot \bo)
\frac{|\bv_{12} \cdot \bo|^{\gamma}}{\Delta v ^{d-1}} P(\bv_2) \delta (\bs+\bo) \delta \left(\Delta v+ \frac{1+\alpha}{2} |\bv_{12} \cdot \bo|\right)\,\,\,,
\end{equation}
where $\Delta v$ and $\bs$ are defined by $\bv_1'-\bv_1 = \Delta v \,\,\bs$. Then, performing the angular 
integration over $\bo$, one obtains:
\begin{equation}
K(\bv_1 \to \bv_1')= \frac{v_0^{1- \gamma}}{l_0} \int \dd \bv_2 \Theta (\bv_{12} \cdot \bs)
\frac{|\bv_{12} \cdot \bs|^{\gamma}}{\Delta v ^{d-1}} P(\bv_2) \delta \left(\Delta v + \frac{1+ \alpha}{2}
|\bv_{12} \cdot \bs|\right) \,\,\,.
\end{equation}
 Denoting by $v_{2 \sigma}$ the component of $\bv_2$ parallel to $\bs$, and by $\bv_{2 \tau}$ the 
 $(d-1)$-dimensional vector in the hyper-plane perpendicular to $\bs$, the above equation is rewritten
 as;
\begin{equation}
K(\bv_1 \to \bv_1')= \frac{v_0^{1- \gamma}}{l_0} \int \dd v_{2 \sigma} \dd \bv_{2 \tau} \Theta (\bv_{12} \cdot \bs) \frac{|\bv_{12} \cdot \bs|^{\gamma}}{\Delta v ^{d-1}} P(v_{2 \sigma}, \bv_{2 \tau}) \delta
\left(\Delta v + \frac{1+ \alpha}{2} |\bv_{12} \cdot \bs|\right)\,\,\,.
\end{equation}
Finally, integrating over $\dd v_{2 \sigma}$, Eq. (\ref{eq:K}) is easily recovered.

\end{document}